\documentclass[twocolumn]{aastex631}

\usepackage{graphicx}	
\usepackage{amsmath}	
\usepackage{subfigure}
\usepackage{natbib}
\usepackage{color}
\usepackage{tabularx}
\usepackage{longtable}
\usepackage{bm}
\usepackage{threeparttable}
\usepackage{enumerate}
\usepackage[normalem]{ulem}
\usepackage{verbatim}

\newcommand{\code}[1]{{\texttt{#1}}}
\newcommand{\tess}{{\it TESS}}
\newcommand{\kepler}{{\it Kepler}}

\newcommand{\gaia}{{\it Gaia}}

\begin{document}

\title{\large Occurrence rate of hot Jupiters around early-type M dwarfs based on TESS data}

\correspondingauthor{Tianjun Gan}
\email{gtj18@mails.tsinghua.edu.cn}

\author[0000-0002-4503-9705]{Tianjun~Gan}
\affil{Department of Astronomy, Tsinghua University, Beijing 100084, People's Republic of China}

\author{Sharon X. Wang}
\affil{Department of Astronomy, Tsinghua University, Beijing 100084, People's Republic of China}

\author{Songhu Wang}
\affil{Department of Astronomy, Indiana University, Bloomington, IN 47405, USA}

\author[0000-0001-8317-2788]{Shude Mao}
\affil{Department of Astronomy, Tsinghua University, Beijing 100084, People's Republic of China}
\affil{National Astronomical Observatories, Chinese Academy of Sciences, 20A Datun Road, Chaoyang District, Beijing 100012, People's Republic of China}

\author[0000-0003-0918-7484]{Chelsea X. Huang}
\affil{University of Southern Queensland, Centre for Astrophysics, West Street, Toowoomba, QLD 4350, Australia}

\author[0000-0001-6588-9574]{Karen A. Collins}
\affil{Center for Astrophysics ${\rm \mid}$ Harvard {\rm \&} Smithsonian, 60 Garden Street, Cambridge, MA 02138, USA}

\author[0000-0002-3481-9052]{Keivan G. Stassun}
\affil{Department of Physics and Astronomy, Vanderbilt University, 6301 Stevenson Center Ln., Nashville, TN 37235, USA}
\affil{Department of Physics, Fisk University, 1000 17th Avenue North, Nashville, TN 37208, USA}

\author[0000-0002-1836-3120]{Avi Shporer}
\affil{Department of Physics and Kavli Institute for Astrophysics and Space Research, Massachusetts Institute of Technology, Cambridge, MA 02139, USA}

\author{Wei Zhu}
\affil{Department of Astronomy, Tsinghua University, Beijing 100084, People's Republic of China}

\author{George~R.~Ricker}
\affil{Department of Physics and Kavli Institute for Astrophysics and Space Research, Massachusetts Institute of Technology, Cambridge, MA 02139, USA}

\author{Roland~Vanderspek}
\affil{Department of Physics and Kavli Institute for Astrophysics and Space Research, Massachusetts Institute of Technology, Cambridge, MA 02139, USA}

\author{David~W.~Latham}
\affil{Center for Astrophysics ${\rm \mid}$ Harvard {\rm \&} Smithsonian, 60 Garden Street, Cambridge, MA 02138, USA}

\author{Sara~Seager}
\affil{Department of Physics and Kavli Institute for Astrophysics and Space Research, Massachusetts Institute of Technology, Cambridge, MA 02139, USA}
\affil{Department of Earth, Atmospheric and Planetary Science, Massachusetts Institute of Technology, 77 Massachusetts Avenue, Cambridge, MA 02139, USA}
\affil{Department of Aeronautics and Astronautics, MIT, 77 Massachusetts Avenue, Cambridge, MA 02139, USA}

\author{Joshua~N.~Winn}
\affil{Department of Astrophysical Sciences, Princeton University, 4 Ivy Lane, Princeton, NJ 08544, USA}

\author{Jon~M.~Jenkins}
\affil{NASA Ames Research Center, Moffett Field, CA 94035, USA}

\author{Khalid Barkaoui}
\affil{Department of Earth, Atmospheric and Planetary Science, Massachusetts Institute of Technology, 77 Massachusetts Avenue, Cambridge, MA 02139, USA}
\affil{Astrobiology Research Unit, Universit\'e de Li\`ege, 19C All\'ee du 6 Ao\^ut, 4000 Li\`ege, Belgium}
\affil{Instituto de Astrof\'isica de Canarias (IAC), Calle V\'ia L\'actea s/n, 38200, La Laguna, Tenerife, Spain}

\author[0000-0003-3469-0989]{Alexander A.\ Belinski}
\affil{Sternberg Astronomical Institute, M.V. Lomonosov Moscow State University, 13, Universitetskij pr., 119234, Moscow, Russia}

\author{David R. Ciardi}
\affil{NASA Exoplanet Science Institute, Caltech/IPAC, Mail Code 100-22, 1200 E. California Blvd., Pasadena, CA 91125, USA}

\author{Phil Evans}
\affil{El Sauce Observatory, Coquimbo, 1870000, Chile}

\author{Eric Girardin}
\affil{Grand Pra Observatory, 1984 Les Hauderes, Switzerland}

\author[0000-0003-4147-5195]{Nataliia A.\ Maslennikova}
\affil{Sternberg Astronomical Institute, M.V. Lomonosov Moscow State University, 13, Universitetskij pr., 119234, Moscow, Russia}
\affiliation{Faculty of Physics, Moscow State University, 1 bldg. 2, Leninskie Gory, Moscow 119991, Russia}

\author{Tsevi Mazeh}
\affil{School of Physics and Astronomy, Tel Aviv University, Tel Aviv, 6997801, Israel}

\author{Aviad Panahi}
\affil{School of Physics and Astronomy, Tel Aviv University, Tel Aviv, 6997801, Israel}

\author{Francisco J. Pozuelos}
\affil{Astrobiology Research Unit, Universit\'e de Li\`ege, 19C All\'ee du 6 Ao\^ut, 4000 Li\`ege, Belgium}
\affil{Space sciences, Technologies and Astrophysics Research (STAR) Institute, Universit\'e de Li\`ege, Belgium}
\affil{Instituto de Astrofísica de Andalucía (IAA-CSIC), Glorieta de la Astronomía s/n, 18008 Granada, Spain}

\author{Don J. Radford}
\affil{American Association of Variable Star Observers, 49 Bay State Road, Cambridge, MA 02138, USA}

\author{Richard P. Schwarz}
\affil{Center for Astrophysics ${\rm \mid}$ Harvard {\rm \&} Smithsonian, 60 Garden Street, Cambridge, MA 02138, USA}

\author[0000-0002-6778-7552]{Joseph D. Twicken}
\affil{NASA Ames Research Center, Moffett Field, CA 94035, USA}
\affil{SETI Institute, 339 Bernardo Ave., Suite 200, Mountain View, CA  94043, USA}

\author{Ana{\"e}l W{\"u}nsche}
\affil{Observatoire des Baronnies Provencales, 05150 Moydans, France}

\author{Shay Zucker}
\affil{School of Physics and Astronomy, Tel Aviv University, Tel Aviv, 6997801, Israel}







\begin{abstract}

We present an estimate of the occurrence rate of hot Jupiters ($7\ R_{\oplus}\leq R_{p}\leq 2\ R_{J}$, $0.8 \leq P_{b}\leq 10$ days) around early-type M dwarfs based on stars observed by TESS during its Primary Mission. We adopt stellar parameters from the TESS Input Catalog, and construct a sample of 60,819 M dwarfs with $10.5 \leq T_{\rm mag}\leq 13.5$, effective temperature $2900 \leq T_{\rm eff}\leq 4000$ K and stellar mass $0.45\leq M_{\ast}\leq 0.65\ M_{\odot}$. We conduct a uninformed transit search using a detection pipeline based on the box least square search and characterize the searching completeness through an injection and recovery experiment. We combine a series of vetting steps including light centroid measurement, odd/even and secondary eclipse analysis, rotation and transit period synchronization tests as well as inspecting the ground-based photometric, spectroscopic and imaging observations. Finally, we find a total of nine planet candidates, all of which are known TESS objects of interest. We obtain an occurrence rate of $0.27\pm0.09\%$ for hot Jupiters around early-type M dwarfs that satisfy our selection criteria. Compared with previous studies, the occurrence rate of hot Jupiters around early-type M dwarfs is smaller than all measurements for FGK stars, although they are consistent within 1--2$\sigma$. There is a trend that the occurrence rate of hot Jupiters has a peak at G dwarfs and falls towards both hotter and cooler stars. Combining results from transit, radial velocity and microlensing surveys, we find that hot Jupiters around early-type M dwarfs possibly show a steeper decrease in occurrence rate per logarithmic semi-major axis bin (${{\rm d}N}/{\rm d}\log_{10} a$) when compared with FGK stars.

\end{abstract}

\keywords{methods: statistical - planetary systems - planets and satellites - stars: low-mass - techniques: photometric}


\section{Introduction}

Even more than a quarter century after the first detection of a hot Jupiter \citep{Mayor1995}, the study of the formation history of giant planets remains a hot topic. The \kepler\ and {\it K2} space missions \citep{Borucki2010,Howell2014} led to the discovery of hundreds of transiting Jupiters, which enabled the studies of the frequency of such planets in our galaxy. \cite{Fressin2013} found that every star surveyed by \kepler\ has an average probability of $0.43\pm0.05\%$ to host a hot Jupiter. Similar occurrence rates of $0.43\pm0.07\%$ and $0.57\pm0.03\%$ were also independently measured by \cite{Masuda2017} and \cite{Petigura2018}. While the results from radial velocity (RV) surveys (e.g., $1.5\pm0.6\%$, \citealt{Cumming2008}; $0.9\pm0.4\%$, \citealt{Mayor2011}; $1.2\pm0.4\%$, \citealt{Wright2012}) are higher than that from transit missions, such difference is suspected to be related to host star properties such as stellar mass and metallicity \citep{Wright2012}. Therefore, grouping mixed stellar samples into different metallicity and mass bins and looking into their Jupiter occurrence rates separately could help probe the formation channel of gas giants and relieve this tension. 

Early works reported that the presence of stars hosting giant planets rises with increasing stellar metallicity \citep{Gonzalez1997,Santos2004,Fischer2005,Sousa2011}, which supports the core accretion planet formation model \citep{Pollack1996}. More recently, \cite{Petigura2018} went a step further and found that the tendency of metal-rich stars to have a higher probability of hosting a giant planet is greater for decreasing orbital period. In terms of stellar mass, \cite{Zhou2019} claimed a weak anti-correlation between occurrence rates of hot Jupiters ($P_{b}\leq 10$ days, where $P_{b}$ is the planet orbital period) and host star mass when splitting the full sample into three stellar types ($0.26\pm0.11\%$ for A stars, $0.43\pm0.15\%$ for F stars, $0.71\pm0.31\%$ for G stars). Recent work from \cite{Beleznay2022} also found a correlation between higher hot Jupiter abundance and lower stellar mass, with hot Jupiter occurrence rates of $0.29\pm0.05\%$, $0.36\pm0.06\%$ and $0.55\pm0.14\%$ for AFG stars, respectively.


Though many studies have been carried out to investigate the occurrence rate of hot Jupiters, most of them focused on AFGK stars. Few relevant studies were extended to the M dwarfs even though M stars are the most abundant stellar population in the Milky Way galaxy \citep{Henry2006}. This bias is mainly a result of rare detections. First, the frequency of such systems may be intrinsically low, as predicted by theoretical works \citep[e.g.,][]{Laughlin2004,Ida2005,Kennedy2008,Liu2019,Burn2021}, due to the low mass as well as the low surface density of protoplanetary disks around M dwarfs. Moreover, the probability of a planet transiting an M dwarf ($p\propto R_{\ast}M_{\ast}^{-1/3} P_{b}^{-2/3}$, $R_{\ast}$ and $M_\ast$ are the stellar radius and mass) is 2--3 times smaller than that for AFGK stars, which leads to a lower detection rate for the same orbital periods. While long-term ground-based transit surveys have made some discoveries (e.g., HATS-6b, \citealt{Hartman2015}; NGTS-1b, \citealt{Bayliss2018}; HATS-71b, \citealt{Bakos2020}; HATS-74Ab and HATS-75b, \citealt{Jordan2022}), these do not represent a homogenous and complete sample due to observational bias. Owing to different environmental conditions, the precision of ground-based photometry cannot stay stable over months and years. This may affect the transit signal search and the final estimation of occurrence rate. Additionally, unlike continuous space observations, ground-based observations are limited by day-night windows, visibility of the stars, as well as technical interruptions, which may create aliasing signals and pose challenges for planet detection and the characterization of search completeness. Finally, the faintness of M dwarfs make it challenging to obtain high signal-to-noise (SNR) spectra and measure precise radial velocity to confirm their planetary nature (e.g., \citealt{Butler2006,Howard2010,Morales2019}). 


\cite{Endl2006} first estimated an upper limit on the frequency of close-in Jovian planets around M dwarfs with semi-major axis $\rm a<1\ AU$ as $<1.27\%$ ($1\sigma$ confidence level) based on RV observations. A similar upper limit result of $1.7-2.0\%$ was also reported by \cite{Kovacs2013} at a $2\sigma$ confidence level for short-period ($0.8\leq P_{b}\leq 10$ days) giant planets around M dwarfs through WFCAM transit surveys. We also refer the readers to \cite{Morton2014}, who did a related study focusing on transiting planets with $R_{p}<4\ R_{\oplus}$, smaller than our lower cutoff of ``giant'' planets, around cool stars in the \kepler\ catalog. With the help of the California Planet Survey \citep{Howard2010}, \cite{Johnson2010} obtained a rate of $3.4^{+2.2}_{-0.9}\%$ that stars with mass below $0.6\ M_{\odot}$ hosting a gas giant with $M_{p}>0.3\ M_{J}$ within 2.5 AU. More recently, \cite{Sabotta2021} reported an occurrence rate upper limit of $3\%$ on hot Jupiters with $100<M_{p}\sin i<1000\ M_{\oplus}$ and $P_{b}<10$ days around M stars through the CARMENES RV survey. Moving outward, the gravitational microlensing technique \citep{Mao1991} is most sensitive to planets at 1--10 AU while the typical host stars of planetary systems discovered through microlensing are M dwarfs. Several statistical studies show that the frequency of microlensing cold Jupiters (planet-to-star mass ratio $> 10^{-3}$) is of the order of 5\% \citep{mufun,Cassan2012,Suzuki2016,Wise}. Based on a combination of long-term RV and high-contrast imaging surveys, \cite{Montet2014} determined that $6.5\pm3.0\%$ M dwarfs harbor a giant planet with mass $1 \leq M_{p}\leq 13\ M_{J}$ located within 20 AU. 

The Transiting Exoplanet Survey Satellite (\tess, \citealt{Ricker2015}), which is performing a nearly all-sky transit survey, opens a new window to enlarge the number of detections of hot Jupiters around M dwarfs. More importantly, \tess\ provides an opportunity to build a homogeneous magnitude-limited M dwarf sample to search for transiting gas giants and estimate their frequency. Additionally, the appearance of new-generation ground-based near-infrared spectroscopic facilities (e.g., HPF, \citealt{Mahadevan2014}; SPIRou, \citealt{Donati2020}) as well as optical instruments on large telescopes (e.g., MAROON-X, \citealt{Seifahrt2018}) enable precise follow-up RV observations for faint M dwarfs and further characterization of the planets around them. There have been several confirmed hot Jupiters around M dwarfs found by \tess\ already (e.g., TOI-530b, \citealt{Gantoi530}; TOI-3629b and TOI-3714b, \citealt{Canas2022}; TOI-3757b, \citealt{Kanodia2022}). 

Here, we present an estimation of the occurrence rate of hot Jupiters (defined as $7\ R_{\oplus}\leq R_{p}\leq 2\ R_{J}$, $0.8 \leq P_{b}\leq 10$ days) around early-type M dwarfs based on the stars observed by \tess\ during the its Primary Mission. We organize the paper as follows: In Section \ref{sample_selection}, we detail how we build our stellar sample. Section \ref{planet_detection_pipeline} describes the detection pipeline we used to uniformly search for planet candidates. The vetting steps and ground-based follow-up observations are presented in Sections \ref{vetting} and \ref{candidate_followup_observation}. We depict the completeness of our detection pipeline through an injection and recovery experiment in Section \ref{injection_recovery} and show the occurrence rate results in Section \ref{occurrence_rate_cell}. We discuss our findings including new planet candidates we identified in Section \ref{discussion} before we conclude in Section \ref{conclusion}.


\section{Sample Selection}\label{sample_selection}
In addition to the pre-selected core planet search stars ($\sim$200,000) that received 2-minute cadence observations, \tess\ also saved the images of its entire field of view every 30 minutes during the TESS 2-year Primary Mission, and every 10 minutes during the Extended Mission \citep{Ricker2015}. After these Full Frame Images (FFIs) were downloaded, they were processed by the MIT Quick-Look Pipeline \citep[QLP;][]{QLP2020a,QLP2020b}. The QLP extracts raw light curves of a magnitude limited ($T_{\rm mag}\leq 13.5$) stellar sample by performing a simple aperture photometry with an optimal-size aperture. The data products of the \tess\ Primary Mission (Sectors 1-26) include 14,773,977 and 9,602,103 light curves from individual sectors for stars in the Southern ecliptic hemisphere (Sectors 1--13) and the Northern ecliptic hemisphere (Sectors 14--26), respectively.\footnote{\url{ https://archive.stsci.edu/hlsp/qlp}} 

We build our stellar sample based on all stars observed during the \tess\ Primary Mission that have QLP light curve for at least one Sector. We combine the target list files of each Sector, which contain the TIC ID, R.A.(J2000, deg) and Dec.(J2000, deg), and remove duplicated entries. We finally find a total of 14,849,252 objects. 

To build a secure M-dwarf sample, we first cross-match our full target list with the \tess\ Input Catalog v8 \citep{Stassun2019tic} through TIC ID and only keep stars that belong to the Cool Dwarf List. The Cool Dwarf List is a sub-sample of the Cool Dwarf Catalog \citep{Muirhead2018}. Basically, it takes the \gaia\ DR2 astrometry \citep{Gaia2018} as well as the broadband photometry information from both \gaia\ and {\it 2MASS} \citep{Cutri:2003,skrutskie2006} into account, and calculates stellar mass $M_{\ast}$ and radii $R_{\ast}$ based on the empirical polynomial relations with absolute $K_{s}$-band magnitude $M_{K}$ \citep{Mann2015,Mann2019}. The precisions of stellar radius and mass estimation are about 2-5\% and 2–3\%, respectively. Effective temperature $T_{\rm eff}$ is computed and calibrated onto observed spectra following the procedure described in \cite{Mann2013b}. The number of mid-to-late type M dwarfs with $T_{\rm mag}$ greater than 18 or $T_{\rm eff}$ less than 2700 K were significantly limited in the Cool Dwarf List \citep[See Figure 16 in][]{Stassun2019tic} due to the parallax measurement signal-to-noise ratio cut ($\rm SNR>5$) and the required $M_{K}$ magnitude criteria ($4.5<M_{K}<10.0$). Second, we remove all stars without distance measurements or distance uncertainties. In this step, we threw out objects that may have problems with the distance (i.e., parallax) determination and only include targets with precise stellar characterization. We next filter out M stars using a conservative effective temperature $T_{\rm eff}$ and stellar mass $M_{\ast}$ cut: $2900 \leq T_{\rm eff}\leq 4000$ K and $0.45\leq M_{\ast}\leq0.65\ M_{\odot}$. We only include early-type M dwarfs in our sample because (1) late type M stars are incomplete in the Cool Dwarf List as aforementioned; (2) the QLP only analyzes stars with $T_{\rm mag}\leq 13.5$ so only a few cool dwarfs have QLP-extracted light curves ready to use.

\begin{figure}
\includegraphics[width=0.49\textwidth]{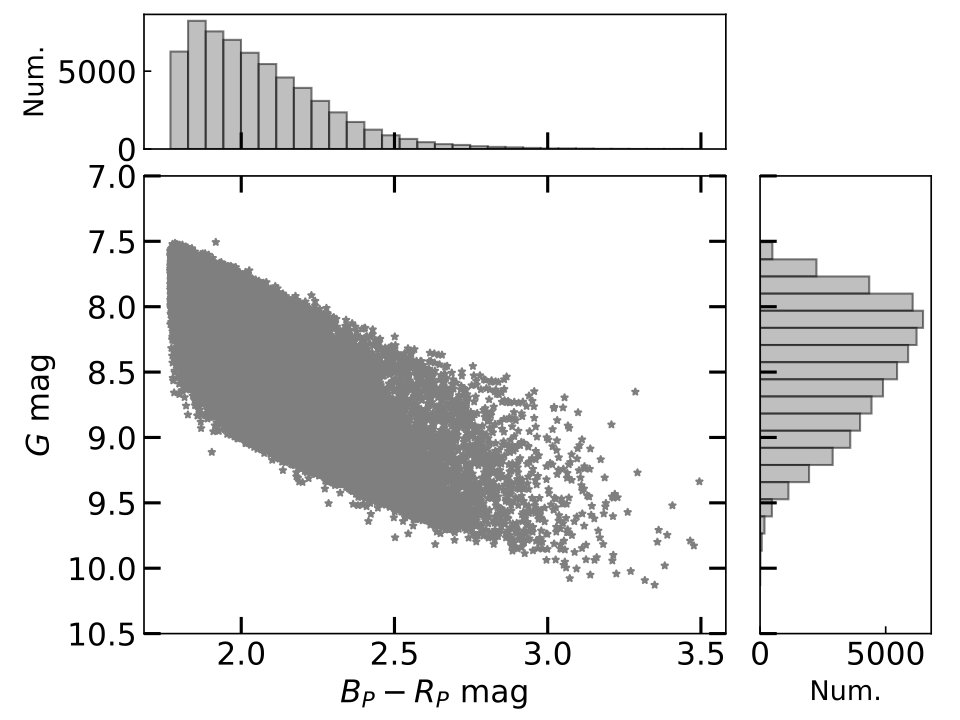}
\caption{\gaia\ color-magnitude diagram of the 60,819 M dwarfs we selected for this study.}
\label{CMD}
\end{figure}
\begin{figure*}
\includegraphics[width=0.95\textwidth]{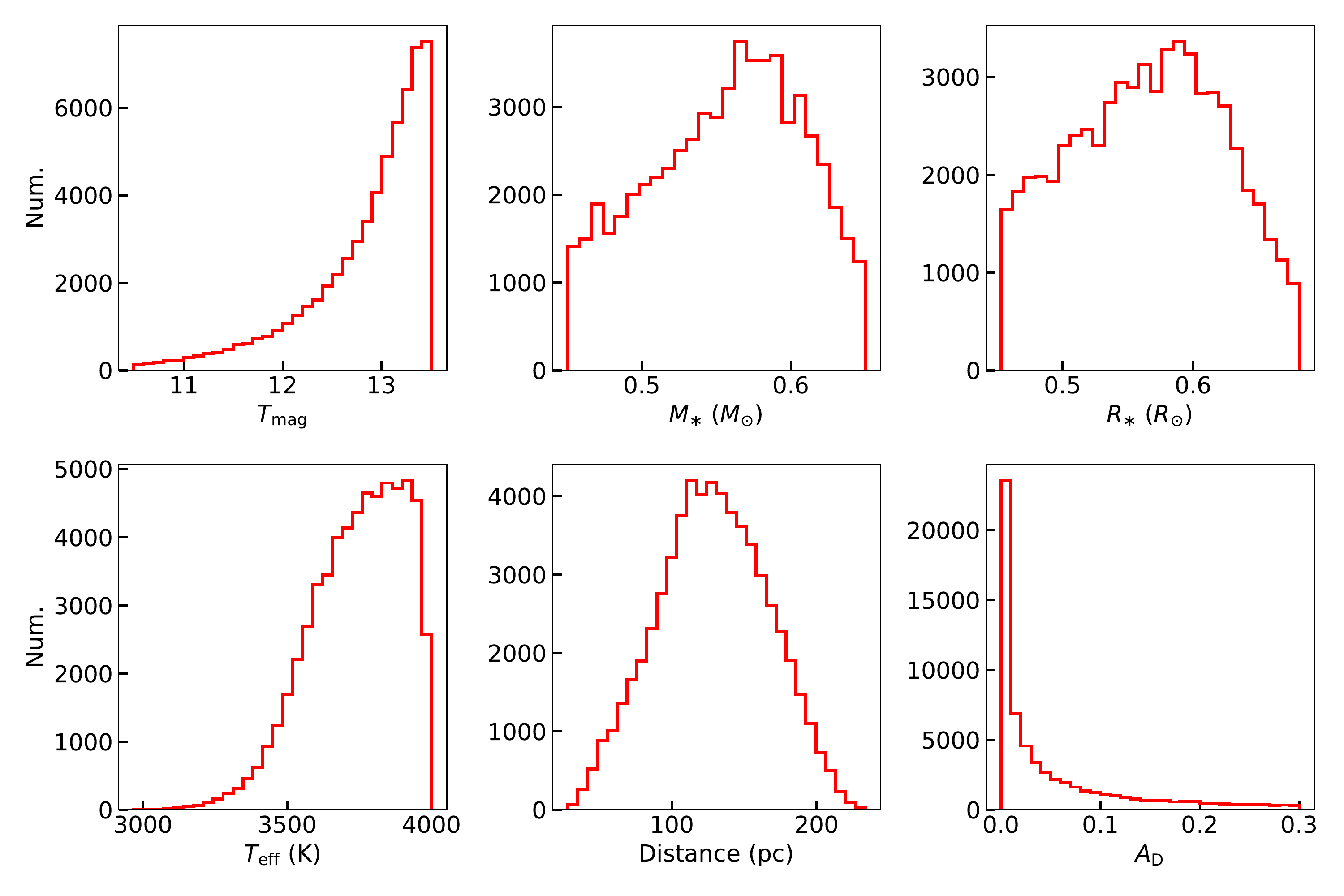}
\caption{Stellar property distribution of our M dwarf sample, including TESS band brightness ($T_{\rm mag}$), mass ($M_{\ast}$), radius ($R_{\ast}$), effective temperature ($T_{\rm eff}$), distance and flux contamination ratio ($A_{\rm D}$). Stellar parameters are retrieved from the TESS Input Catalog \citep{Stassun2019tic}.}
\label{sample_properties}
\end{figure*}

Finally, we restrict a brightness-limited sample by including objects with $10.5 \leq T_{\rm mag}\leq 13.5$ and remove stars with dilution factors greater than 0.3. Since \tess\ has a large pixel scale ($21\arcsec$/pixel), the ``third-light'' flux provided by bright nearby stars can lead to an underestimated planetary radius \citep{Ciardi2015}, especially for planets around faint M dwarfs. Additionally, the contamination flux will possibly result in incorrect star properties \citep{Furlan2017b,Furlan2020}. The flux contamination ($A_{\rm D}$) reported by TIC v8 is computed as the ratio of total contaminant flux within a radius that depends on the target's brightness to the target star flux \citep{Stassun2017tic,Stassun2019tic}. We conservatively exclude targets with significant dilution ($A_{\rm D}>0.3$), making stars in our final sample relatively isolated and having accurate constraints on stellar properties. We note that we consider the dilution effect and apply this correction factor $A_{\rm D}$ in the transit fit as well as the injection and recovery section (See Sections \ref{vetting} and \ref{injection_recovery}). A total of 60,819 stars pass the above selection function and remain in our sample. We present their color–magnitude diagram in Figure \ref{CMD}. Figure \ref{sample_properties} shows the distribution of stellar properties for our final selected sample. The median uncertainties on mass, radius, effective temperature and distance are $0.02\ M_{\odot}$, $0.02\ R_{\odot}$, 157 K and 0.5 pc, respectively.

\section{Planet Detection}\label{planet_detection_pipeline}
\subsection{Light Curve Pre-processing}\label{light_curve_preprocessing}
\begin{figure*}
\centering
\includegraphics[width=0.95\textwidth]{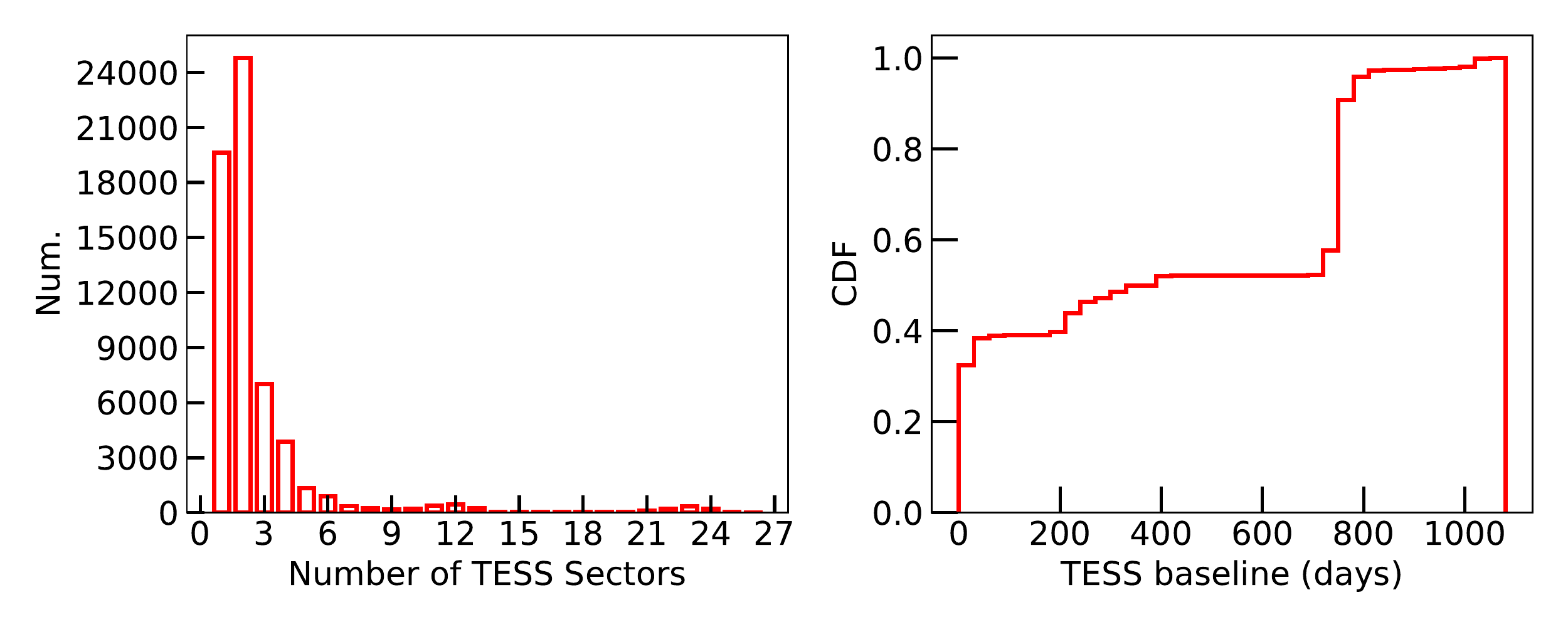}
\caption{{\it Left panel}: Distribution of the number of TESS sectors that have light curves for stars in our stellar sample. {\it Right panel}: The cumulative distribution function of the TESS baseline length of our stellar sample. Many stars were only observed for only a single sector per year while several other stars were observed for a few sectors during different years. The enhancement around 800 days is caused by the revisit of TESS during its Extended Mission.}
\label{TESS_baseline}
\end{figure*}

In order to obtain a high SNR transit detection and better understand the architecture of each system, we make use of all available QLP light curves of our stellar sample from both \tess\ Primary and Extended Mission. We retrieve the light curves of each target from Mikulski Archive for Space Telescopes\ (MAST\footnote{\url{http://archive.stsci.edu/tess/}}) via \code{astroquery} \citep{Ginsburg2019}. To improve the precision of light curves, we ignore entries where the quality flag is assigned non-zero, which indicates anomalies in the data or images \citep{QLP2020a}. Despite this, the raw light curves of most stars still have a few data points with abnormally high flux values. We thus calculate the 99.5th percentile of each light curve and exclude 0.5\% points with the highest flux for all stars in our sample, which might be related to instrumental or systematic noise. We show the \tess\ baseline length distribution of our final stellar sample in Figure \ref{TESS_baseline}.

Second, we perform a uniform detrending by fitting basis spline models for each light curve. During the construction of spline models, we use a running three sigma-clipped median filter. We divide the full light curve into several bins with a binning size of 0.3 days. Within each bin, we calculate the median flux after removing $3\sigma$ outliers. Next, we interpolate the 0.3-day binned full spline that we obtained onto the full observation time stamps with a cubic interpolator. We finally produce the detrended light curve by dividing the original light curve by this interpolated spline function. We use these detrended light curves for candidate search.







\subsection{Candidate Search}\label{Candidate Search}

Our planet detection pipeline is mainly based on the Box Least Square (BLS; \citealt{Kovacs2002}) algorithm\footnote{\url{https://docs.astropy.org/en/stable/timeseries/bls.html}}. Following the methodology described in \cite{Dressing2015}, we first perform a low-resolution transit search. We explore 1,000 uniformly spaced period grids between 0.8 and 10 days. To determine the best duration searching grid, we randomly select 10,000 stars from our sample, generating arbitrary physical parameters with period $P$ between 0.8 and 10 days, impact parameter $b$ below 0.9 and planet size between $7\ R_{\oplus}$ and $2\ R_{J}$, and compute the transit duration assuming a circular orbit \citep{Seager2003}. We find that more than 99\% durations are located between 0.05 and 0.17 days. Consequently, to be conservative, we conduct our search for 10 uniformly spaced transit durations between 0.02 and 0.2 days, where all of our simulated duration values are located in. 

We first compute a BLS periodogram for each detrended light curve. To ensure a relatively clean sample without too many false positives, we require the selected candidates to have a BLS reported maximum SNR ($SNR_{\rm transit}$) greater than 10. Since we use a relatively sparse spline model to detrend the light curves, short-timescale sinusoidal-like stellar variations may be left in the data and cause false alarms. However, we do not expect a strong periodic brightening effect with a similar amplitude comparable with the dimming signal for real transit events. Therefore, we conduct an anti-transit search by constructing another BLS periodogram for the flipped light curve \citep{Wang2014} to identify stellar variability. Similarly, we record the BLS maximum SNR of the anti-transit ($SNR_{\rm anti-transit}$). We define an SNR ratio ($SNR_{\rm transit}/SNR_{\rm anti-transit}$) as an indicator that reflects the robustness of a real transit detection, and we only keep candidates with an SNR ratio greater than 1.5. 

If a candidate passes the above low-resolution transit search above, we next refine the period $P$ and mid-transit timing $T_{0}$ by performing high-resolution BLS runs to examine alias signals. Starting with the period $P_{\rm raw}$ found in the previous step, we calculate all aliasing periods $P_{\rm alias}$ as $ P_{\rm raw}/N$ and $P_{\rm raw}\times N$, where $N$ is a positive integer, and save period values between 0.4 and 12 days. During the high-resolution runs, we focus on the period range of [$P_{\rm alias}-0.1$, $P_{\rm alias}+0.1$] with 1,000 intervals and the same transit duration grid as in the previous search, and loop the BLS search for all aliasing periods. We regard the $P_{\rm BLS}$-$T_{\rm 0,BLS}$-duration set with the highest BLS SNR as the final transit ephemeris. 

Eclipsing binary systems generally have a significant secondary dip, manifesting as a depth difference between odd and even transits. To the contrary, we expect identical odd/even depths for real planetary signatures. To reject such astrophysical false positives, we compare the odd and even depths ($\delta_{\rm odd}$ and $\delta_{\rm even}$) reported by the BLS algorithm. We calculate the odd/even depth difference $\Delta=|\delta_{\rm odd}-\delta_{\rm even}|$ and require all candidates to have $\delta_{\rm odd}/ \Delta$ and $\delta_{\rm even}/ \Delta$ greater than 3. This conservative threshold is set based on a test on several selected binary systems. A more careful investigation of the odd/even difference is performed in Section \ref{secondary_odd_even}.

\begin{figure}
\includegraphics[width=0.5\textwidth]{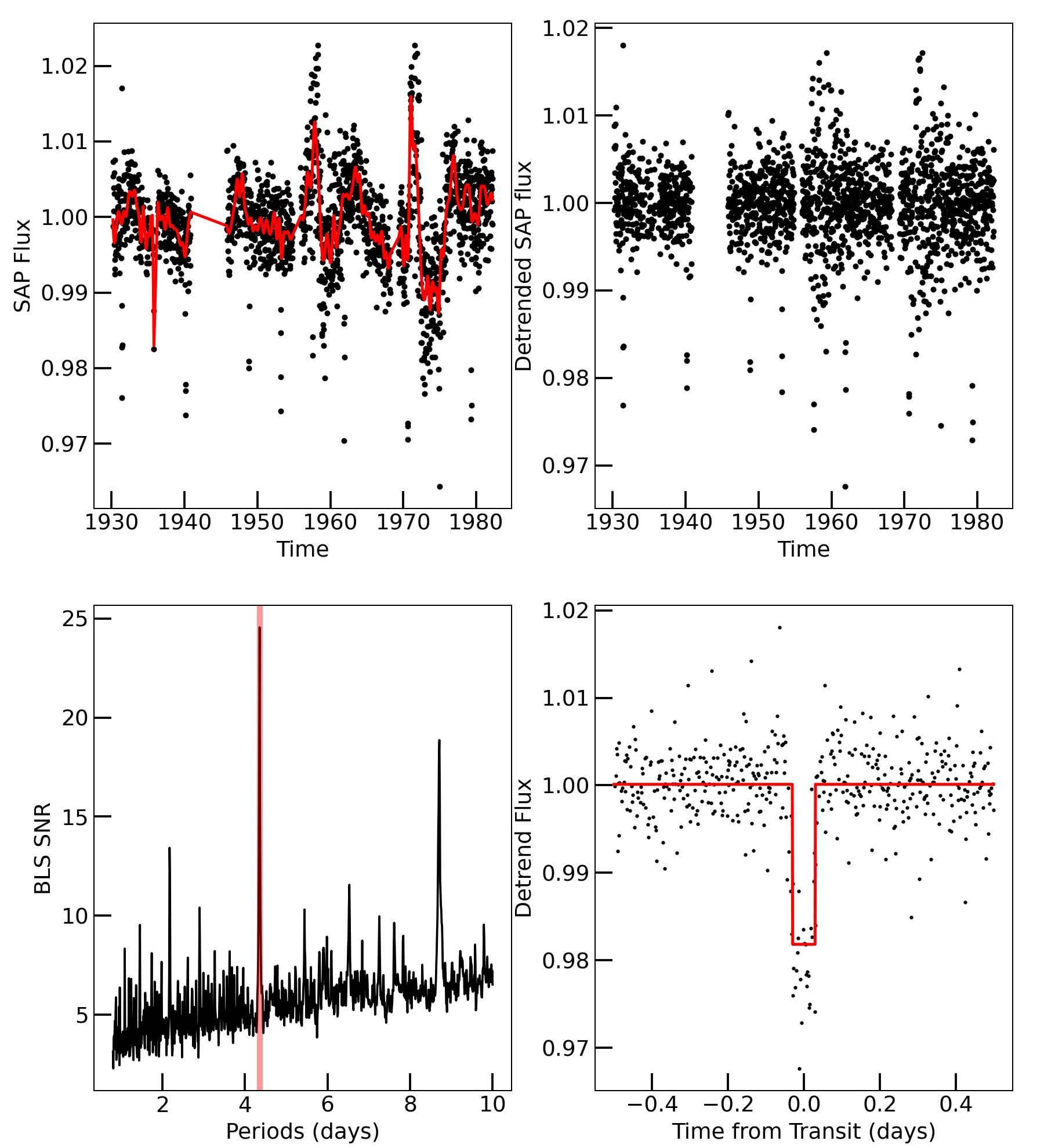}
\caption{An example diagnostic plot of our transit search for TIC 20182780 (TOI-3984). {\it Top panels}: The raw and detrended QLP light curves. The cubic spline model (binning size=0.3 days) used for detrending is shown as a red solid curve. {\it Bottom left panel}: The BLS periodogram of the low-resolution transit search. The detected best period is indicated by a vertical red shaded region. {\it Bottom right panel}: The detrended light curve phase-folded at the final best period found in the high-resolution transit search. The red solid line represents the best BLS model. }
\label{BLS_search}
\end{figure}

Our detection pipeline alerts a total of 437 events in the end. For each candidate, a diagnostic plot is generated as in Figure \ref{BLS_search}, which includes the raw QLP photometry, spline model detrended data, BLS periodogram of the low-resolution search as well as the phase-folded light curve to the transit ephemeris found in the high-resolution search. We note that (1) our detection pipeline only examines the transit-like signal with the highest SNR, and it may miss giant planets around young M dwarfs with strong short-timescale variations. Such incompleteness will be characterized by the injection-recovery test; (2) we do not deal with multi-planet cases in this work as M dwarf systems with close-in hot Jupiters and additional planets are rare, which have not been detected so far.



\subsection{Known TOIs Missed by the Detection Pipeline}

\begin{figure}
\includegraphics[width=0.49\textwidth]{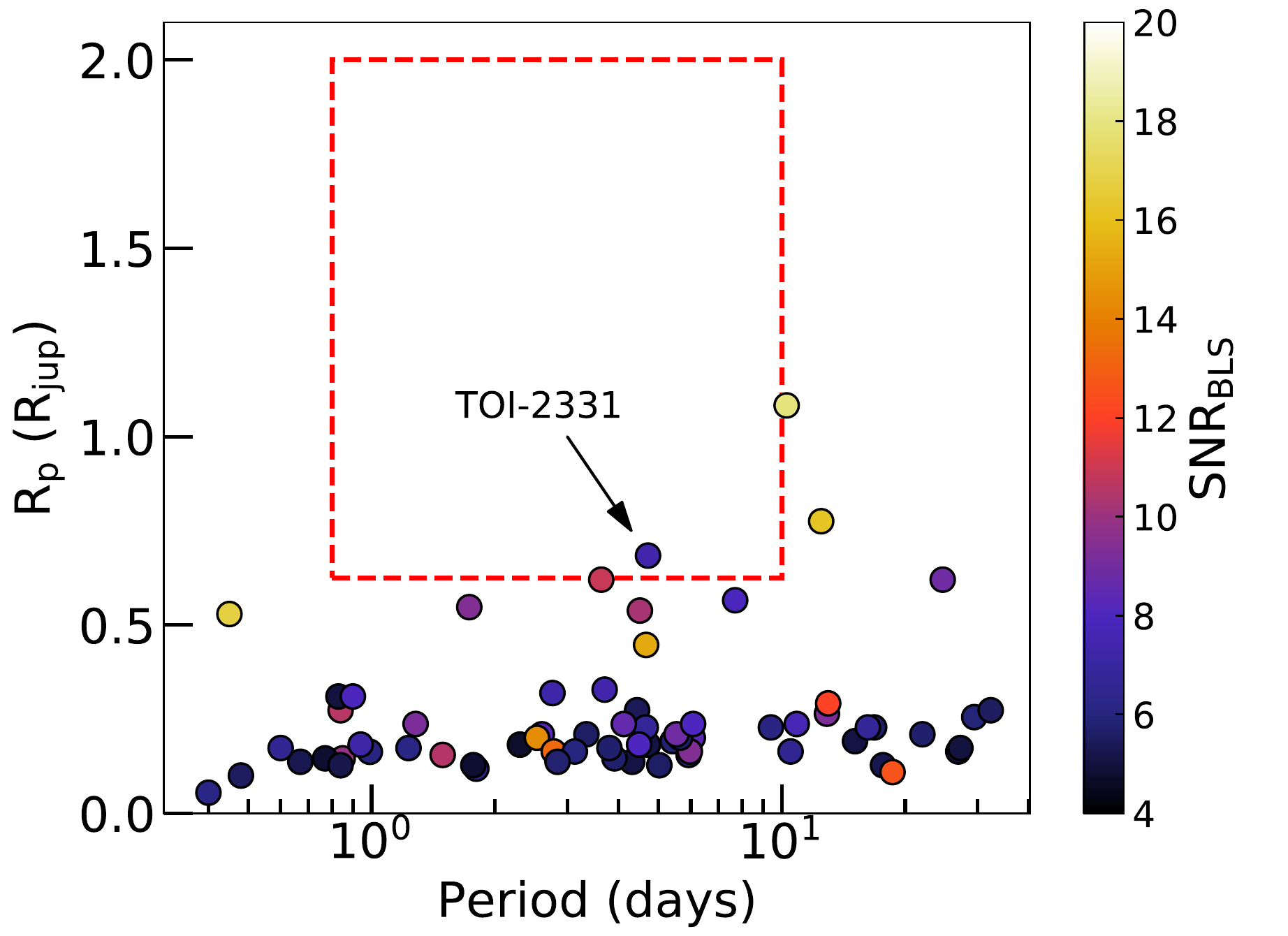}
\caption{The radius-period diagram of 67 TOIs that are in our parent sample but did not trigger an alert by our transit search pipeline. Different colors correspond to different maximum SNR of the low-resolution BLS search. The region within the red dashed box is the parameter space we have searched in this work. Only one target (TOI-2331) in this area was not alerted by our pipeline because its SNR does not satisfy our threshold ($\rm SNR_{BLS}\geq 10$).}
\label{missed_TOI_plot}
\end{figure}

Besides for the 437 candidates studied in this work, there are also known TOIs that were missed by our detection pipeline. We match the catalog of 60,382 stars without a detection with a list of known TOIs, and we found a total of 67 TOIs that were not alerted. We present the full catalog of missed TOIs and report our search results in Table \ref{missed_TOIs}. Figure \ref{missed_TOI_plot} shows their period-radius distribution. Most of these candidates were found in the 2-min cadence light curves extracted by TESS Science Processing Operations Center \citep[SPOC;][]{Jenkins2016}. Given their small companion size around 1--4 $R_{\oplus}$ and short duration time, the transit depth will be diluted in long cadence FFI data. Thus, the BLS SNRs of these small planet candidates are generally low. Among all of these missed candidates, we note that there is one target (TIC 168751223/TOI-2331) that is within the parameter space we have searched in this work. It was not alerted by our detection pipeline because the SNR of our BLS search does not match our minimum threshold (SNR=10). Further ground-based follow up observations ruled out this candidate as by confirming that the eclipse signal is from a nearby binary star system on TIC 168751224 ($\Delta T=3.1$ mag) at 7$\arcsec$ away. 

\section{Vetting}\label{vetting}

\begin{figure}
\includegraphics[width=0.49\textwidth]{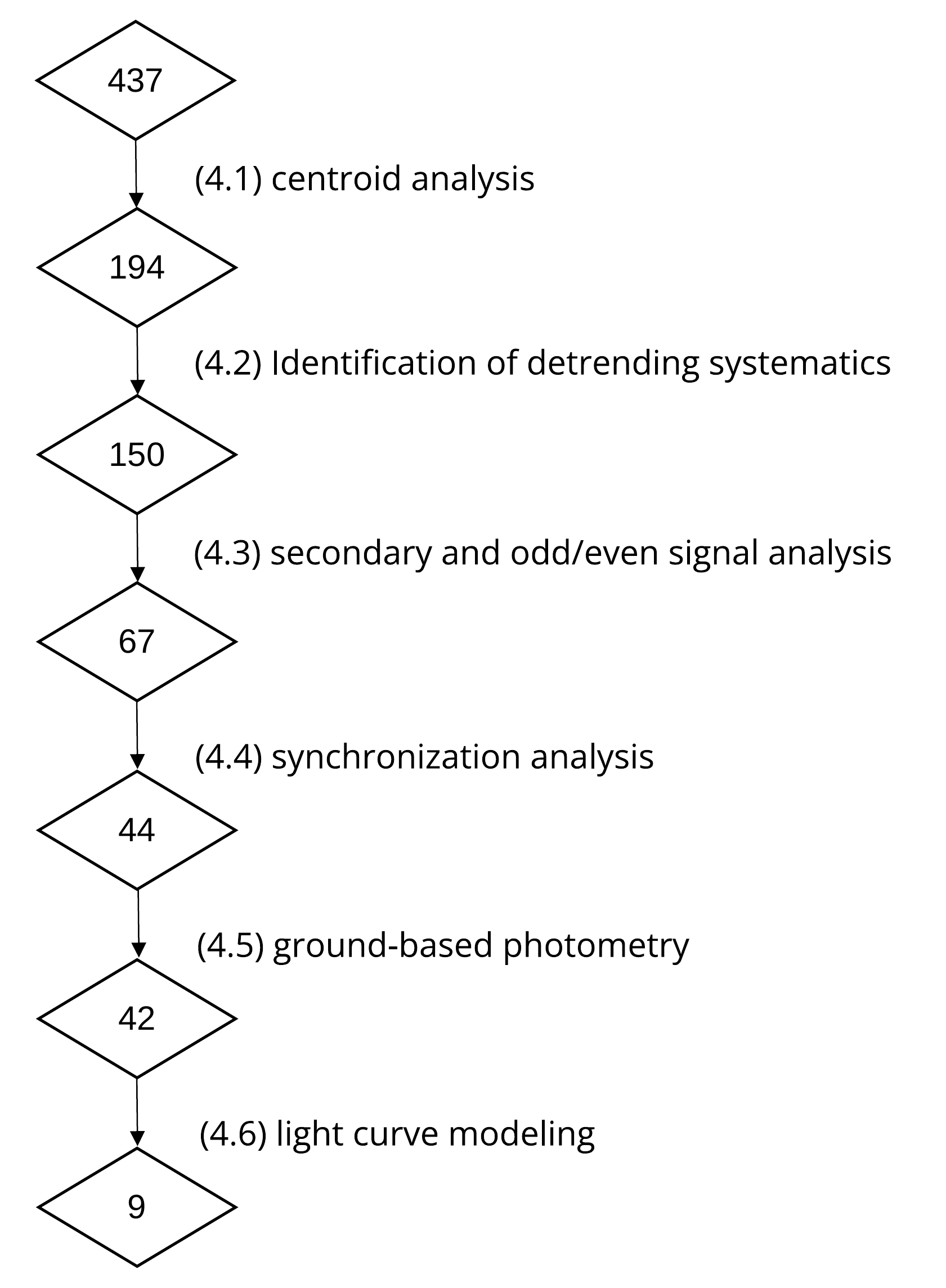}
\caption{A summary of our candidate vetting process. Each box shows the number of remaining candidates after each step. All of these steps are described in Section \ref{vetting}.}
\label{vetting_process}
\end{figure}

We conduct a series of vetting analyses to remove false positives among the 437 candidates found by our detection pipeline. A brief summary of our vetting process is shown in Figure \ref{vetting_process}. 

\subsection{Centroid Analysis}
The QLP produces light curves of each source using several apertures, and identifies an optimal light curve based on the target brightness. Generally, the size of this optimal aperture is larger than 2 \tess\ pixels for M dwarfs with $10.5\leq T_{\rm mag}\leq 13.5$ so the light from nearby eclipsing binaries within $1\arcmin$ may pollute the aperture and cause transit-like signals on the target light curve. We reject such scenarios using the difference image technique \citep{Bryson2013}. We perform a pixel-level centroid offset analysis in the difference image of each candidate with \code{TESS-plots\footnote{\url{https://github.com/mkunimoto/TESS-plots}}} \citep{Kunimoto2022}. \code{TESS-plots} downloads $20\times 20$ pixel cutout of \tess\ FFIs obtained in a certain sector, generates a difference image based on the flux of in- and out-of-transit images and calculates the SNR of each pixel. The light centroid of the difference image should be located around the source position in the direct image (i.e., FFI) if the signal is on target. Otherwise, the nearby eclipsing binary scenario is favored when a large centroid shift happens. We compute a SNR-weighted light centroid ($x_{\rm c}, y_{\rm c}$) in a $7\times7$ pixel difference image centered on target for every candidate using:
\begin{align}
    x_{\rm c} = \frac{\sum_{i=1}^{7}\sum_{j=1}^{7} {\rm SNR}^2_{x_i,y_j} \times x_i}{\sum_{i=1}^{7}\sum_{j=1}^{7} {\rm SNR}^2_{x_i,y_j}}, \\
    y_{\rm c} = \frac{\sum_{i=1}^{7}\sum_{j=1}^{7} {\rm SNR}^2_{x_i,y_j} \times y_j}{\sum_{i=1}^{7}\sum_{j=1}^{7} {\rm SNR}^2_{x_i,y_j}},
\end{align}
where $x_i$ and $y_j$ are the pixel indices, ${\rm SNR}_{x_i,y_j}$ represents the signal-to-noise of pixel ($x_i, y_j$) in the zoomed difference image. 

\begin{figure*}
\subfigure[TIC 14081980 ($d_{\rm c}$=1.2 pixels)]{\includegraphics[width=0.498\textwidth]{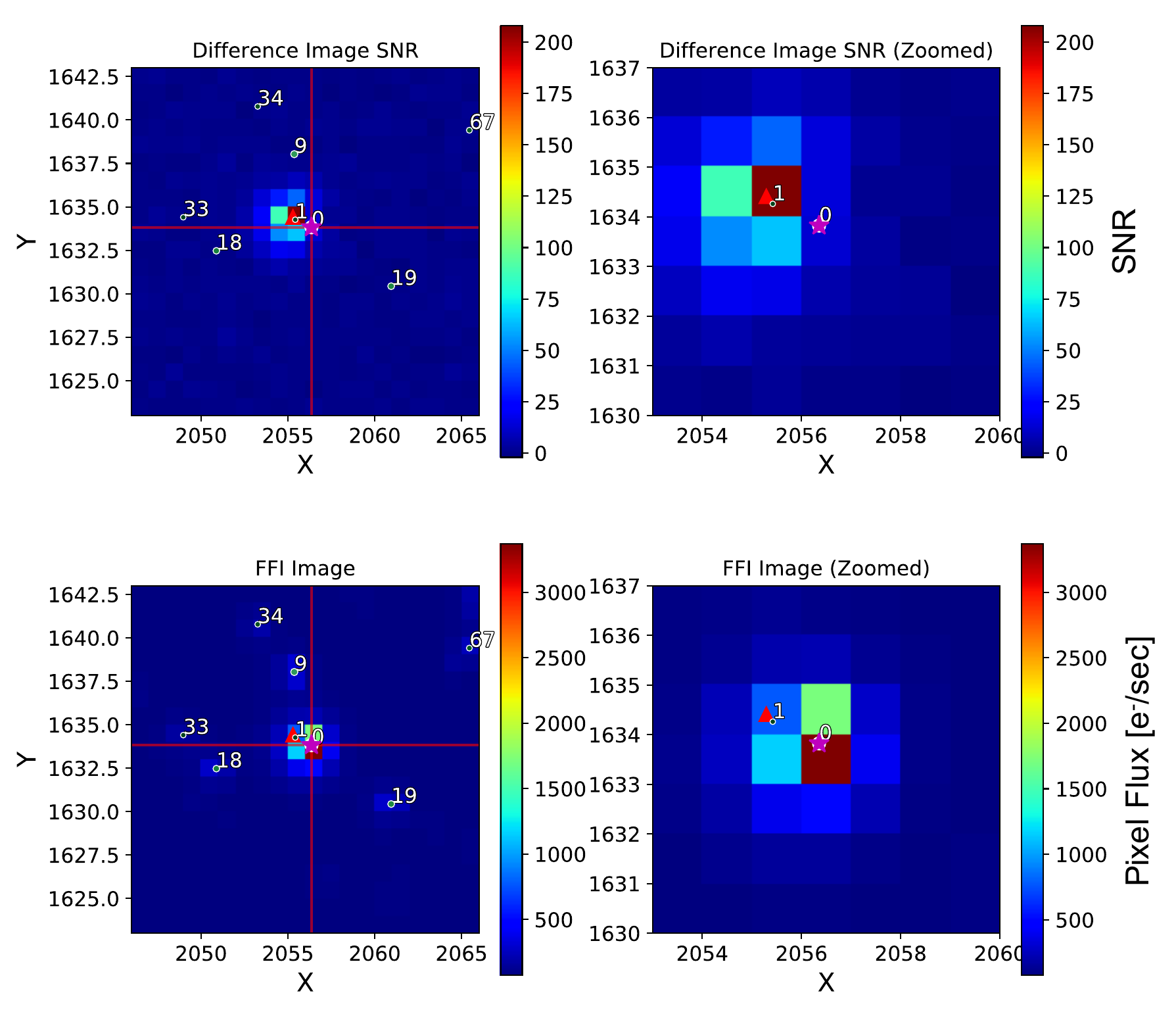}}
\subfigure[TIC 470988013 ($d_{\rm c}$=0.9 pixels)]{\includegraphics[width=0.49\textwidth]{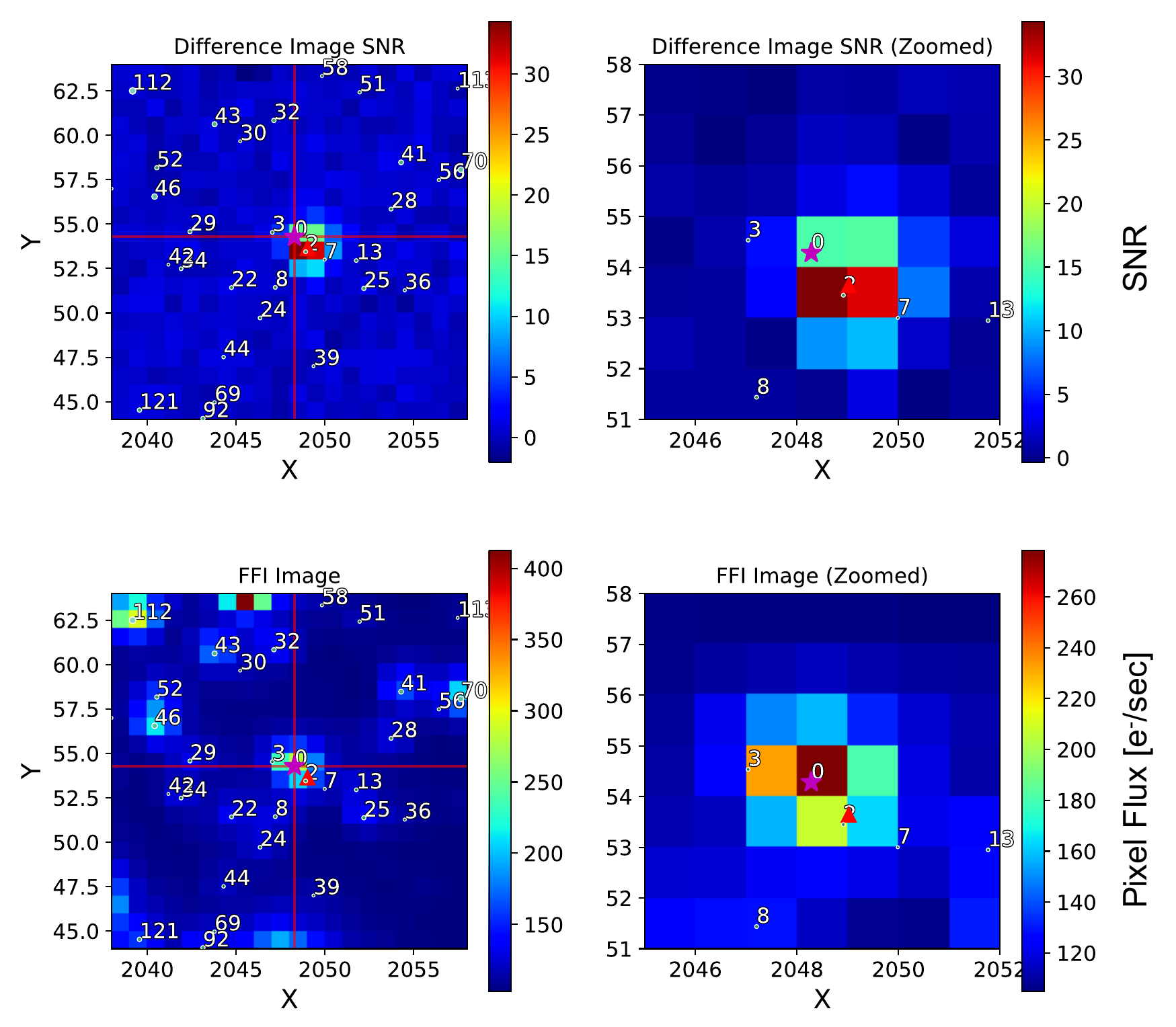}}
\caption{Two example diagnostic plots of our light centroid analysis. TIC 14081980 (left sub figure) is exclude according to the $1$ pixel centroid shift cut while TIC 470988013 (right sub figure) is exclude through visual inspection. {\it Top panels}: A $20\times20$ TESS pixel difference SNR image of TIC 14081980 (left) and a zoomed-in plot of the central $7\times 7$ pixels (right). The target is shown as the magenta star in the center of both images. Nearby stars fainter than the target with $\Delta T\leq4$ mag are plotted as circles. The red triangle represents the SNR-weighted light centroid we measured (see Section \ref{vetting}). {\it Bottom panels}: Similar as above but here are the direct TESS images (FFIs) during out-of-transit. }
\label{centroid_plot}
\end{figure*}

We compare the target position on the difference image with the measured light centroid in the FFI and calculate the centroid shift ($d_{\rm c}$). We generate difference images from different sectors for each star and accept the cases where the light centroid shift of the highest SNR difference image is smaller than 1 pixel. An example, TIC 14081980 with $d_{\rm c}=1.2 $ pixels, that we excluded in this step is shown in the left panel of Figure \ref{centroid_plot}. In total, there are 238 candidates survive after this step. For confirmed planets detected by \tess\ (Table \ref{final_candidates}), all of their centroid shifts are $< 0.5$ pixels. The 1-pixel ($\sim$21$\arcsec$) centroid offset is a conservative threshold to rule out signals from nearby binary or planetary systems, which is also previously used in alerting TOIs by TESS teams \citep{Guerrero2021,Kunimoto2022}. In the SPOC validation reports \citep{Twicken2018}, a $2.5\arcsec$ error term is added in quadrature to the propagated uncertainty in the difference image centroid offsets. The $3\sigma$ centroid offset level for single sector observations is roughly $7.5\arcsec$ (0.35 TESS pixel) for a majority of target stars. This is much smaller than the 1-pixel choice here. The SPOC centroid shifts of confirmed or known planets alerted with transit depths between 5000 to 25000 ppm are smaller than $1\sigma$ ($<3\arcsec$). Excluding candidates with centroid shifts larger than 1 pixel ($\sim$21$\arcsec$) thus give a completeness much higher than 99.7\% so the false negative rate of this step is negligible for our study. Although \cite{Twicken2018} pointed out that the light centroid determination is sometimes unreliable for targets within crowded fields as nearby bright stars will have an effect, we note that we excluded M stars with dilution $A_{\rm D}>0.3$ (see Section \ref{sample_selection}). Therefore, the targets in our sample are relatively isolated and have, in principle, precise light centroid measurements.

\begin{figure}
\includegraphics[width=0.49\textwidth]{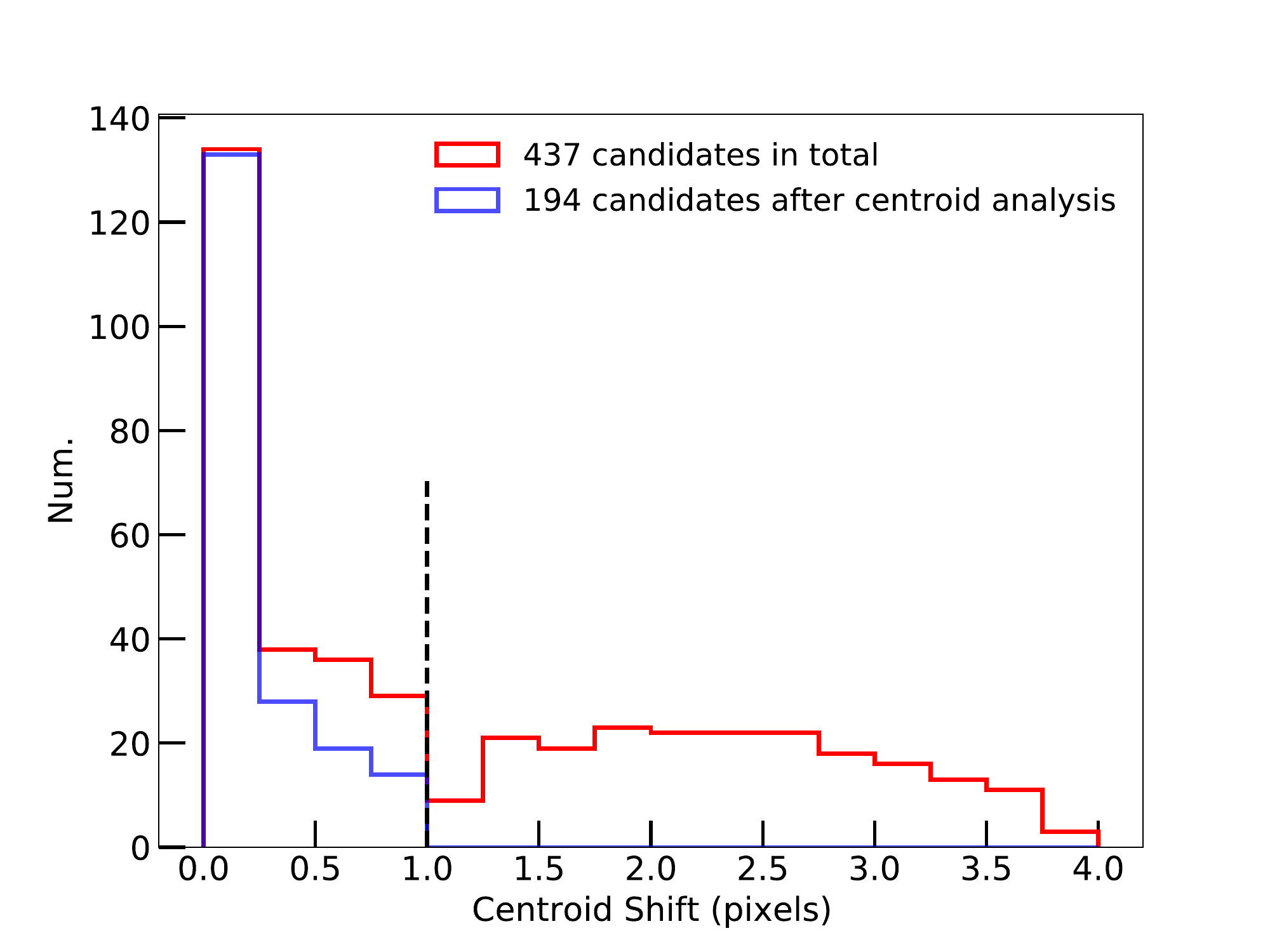}
\caption{The red histogram is the centroid shift distribution of all 437 candidates we found. We filter out targets with centroid shift beyond 1 pixel (vertical black dashed line) as they are certainly nearby eclipsing binaries. The blue histogram shows the distribution of 194 candidates after removing objects whose centroids shift to nearby pixels (see Section \ref{vetting}).}
\label{centroid_distribution}
\end{figure}

Next, we further exclude 44 events with centroid shifting to the same nearby star in an adjacent pixel but with shifts smaller than 1 \tess\ pixel, and the difference images from different \tess\ Sectors give consistent results. We show the example difference images and FFIs of a target TIC 470988013 in our candidate list that we removed with $d_{\rm c}=0.9$ pixels in the right panel of Figure \ref{centroid_plot}. All the other 43 targets show a similar degree of centroid shifts like this object. We note that we keep negligible false negative rates during this step and only exclude obvious nearby eclipsing binary systems. Figure \ref{centroid_distribution} displays the centroid shift distribution of all candidates as well as the 194 remaining candidate events. 

\subsection{Identification of Detrending Systematics}\label{visual_inspection}
We next remove 44 false positives through visual inspection. Among them, the alerts of 11 candidates are caused by systematics and they show an apparent flux drop after detrending. Most of these false signals happen at the edges of \tess\ light curve gaps where the flux changes sharply due to instrumental systematics. Such signals neither show transit-like shape nor appear periodically in the light curves. A total of 31 candidates are alerted due to stars with residual stellar variations after detrending. Our uniform basis spline model failed to fully remove the stellar variability and caused the alarms. We also removed two binary systems (TIC 334790937 and TIC 446963308) with orbital period larger than 10 days and deep primary transit. We show the light curves of all these excluded candidates in Figure \ref{detrending_issues}. 

We note that the exact false negative rate of this step has a negligible effect on our final result, so in our calculation of the occurrence rate, we assume that the false negative rate (or detection completeness) for this sample of 44 stars is the same as the whole sample of $\sim$60k stars. We justify this assumption as follows: First, the false negative rate of this sample of 44 stars is probably not significantly higher than the whole sample (although likely a bit higher due to the increased systematics or stellar variations). We have carefully examined the light curves of these candidates to minimize the possibility of missing bona fide planets. If there are additional periodic transit signals with depth about $1\%$ or larger in the variability or systematics, they would have been picked out easily by eye. Second, even if the false negative rate of this sample is as high as, for example, a few times higher than the whole sample, it would still have an negligible effect on our final statistics. The sample size of 44 is very small compared to the size of the whole sample of $\sim$60k, so the additional uncertainty in the false negative rate of this sample is negligible in the equation calculating the occurrence rate, as it is a negligible fraction in the denominator of equation \ref{eqn:occ-rate}. Therefore, we conclude that these 44 stars will not affect the final statistics.


\subsection{Secondary and Odd/Even Signal Analysis}\label{secondary_odd_even}
Next, we examine the secondary eclipse signals more carefully to identify additional false positives in the candidates. Though we have already placed a constraint on the odd/even transit depth in the detection pipeline, BLS only reports the depths for two models where the period is twice the fiducial period (odd transit, $2P_{\rm BLS},\ T_{\rm 0, BLS}$) and the same period but the phase is offset by one fiducial period (even transit, $2P_{\rm BLS},\ T_{\rm 0, BLS}+P_{\rm BLS}$). In this case, we note that: (1) Candidates with secondary eclipses happen at the fiducial period but have a half fiducial period phase shift ($P_{\rm BLS},\ T_{\rm 0, BLS}+P_{\rm BLS}/2$) will be missed.\footnote{This happens when the candidate has a circular orbit. For eccentric orbits, the center of the secondary eclipse would have a shift.} (2) Our detection pipeline cannot handle cases when the odd/even depths are close to each other but different (i.e., do not satisfy the criteria we set in Section \ref{Candidate Search}: either $\delta_{\rm odd}/ \Delta$ or $\delta_{\rm even}/ \Delta$ smaller than 3). Therefore, we investigate the phase-folded light curves (see Figure \ref{odd_even_secondary}) at a suite of transit ephemerides, i.e., $P_{\rm BLS}, T_{\rm 0, BLS}$; $P_{\rm BLS},\ T_{\rm 0, BLS}+P_{\rm BLS}/2$; $2P_{\rm BLS},\ T_{\rm 0, BLS}$ and $2P_{\rm BLS},\ T_{\rm 0, BLS}+P_{\rm BLS}$. We exclude 83 targets with significant secondary eclipses ($>$ 1\%) or the difference between odd and even depth ($|\delta_{\rm odd}-\delta_{\rm even}|$) is higher than 1\% in this step. 

We note that our odd/even vetting is unlikely to reject real planets with bona fide secondary eclipse signals. Since we set a very conservative threshold on the secondary eclipse ($>1\%$) and the odd/even difference ($>1\%$), an imperfect detrending is unlikely to cause such a large difference between different transits, which is significant and comparable with the transit depth. We visually checked the light curves of these 83 targets and confirmed that the depth differences are astrophysical instead of systematics due to detrending issues. The photometric noise could not cause false odd/even or secondary eclipse signals as deep as 1\% given the photometric precision of these 83 targets, which is all much better than 1\%. Some ultra-short period hot Jupiters have detected secondary signals (e.g., WASP-18b, \citealt{Shporer2019}; TOI-2109b, \citealt{Wong2021}). However, such secondary signals \citep{Twicken2018} would be buried in the noise of the QLP light curves of our M dwarf sample -- planets around M dwarf are much less irradiated by their host star compared to Sun-like stars, leading to lower equilibrium temperature $T_{\rm eq}$ and a shallower secondary eclipse in the optical ($<$ 1 mmag) band. Therefore, our $1\% $ secondary depth cut above is much larger than the expected $<$ 1 mmag signal and it will keep all real planets in this step, meaning a negligible false negative rate. Moreover, the typical standard deviation of the QLP light curves of our sample is around 3--4 mmag, which is insufficient to detect the secondary eclipse signal in our cases.

\subsection{Synchronization Analysis}
Another way of identifying false positive signals is to compare the stellar variation periodicity with the transit signal's periodicity.  Candidates with eclipse signals synchronized with out-of-transit phase variation are unlikely to be real planetary systems, because it is rare to have the stellar rotation period synchronized with the planet orbital period especially for M dwarfs. Based on the empirical relation derived by \cite{Engle2018}, we estimate that early-type M dwarfs with rotation periods smaller than 10 days would have ages below 0.9 Gyr. However, the expected time for a planet to enforce its host star to spin at the same period with the planet's orbital period is much longer than a Hubble time \citep{Zahn1977}. Assuming a $0.5\ M_{J}$ hot Jupiter around a typical early-type M dwarf with a mass and radius of $0.5\ M_{\odot}$ and $0.5\ R_{\odot}$, the synchronization timescale can be approximated by
\begin{equation}
    \tau_{\rm sync} \sim q^{-2}\left(\frac{a}{R_\ast}\right)^{6} {\rm yr} \sim10^{5}\left(\frac{a}{0.05\ AU}\right)^{6}\ {\rm Gyr},
\end{equation}
where $q=M_{p}/M_{\ast}$ is the mass ratio between planet and star, $a/R_{\ast}$ is the planet orbital semi-major axis in units of stellar radius. This is much longer than any astrophysical timescale, hence the correlation between the rotation modulation and the eclipse signal is likely due to ellipsoidal variations caused by tidal distortions and gravity brightening in stellar binaries in these cases. 

\begin{figure}
\includegraphics[width=0.49\textwidth]{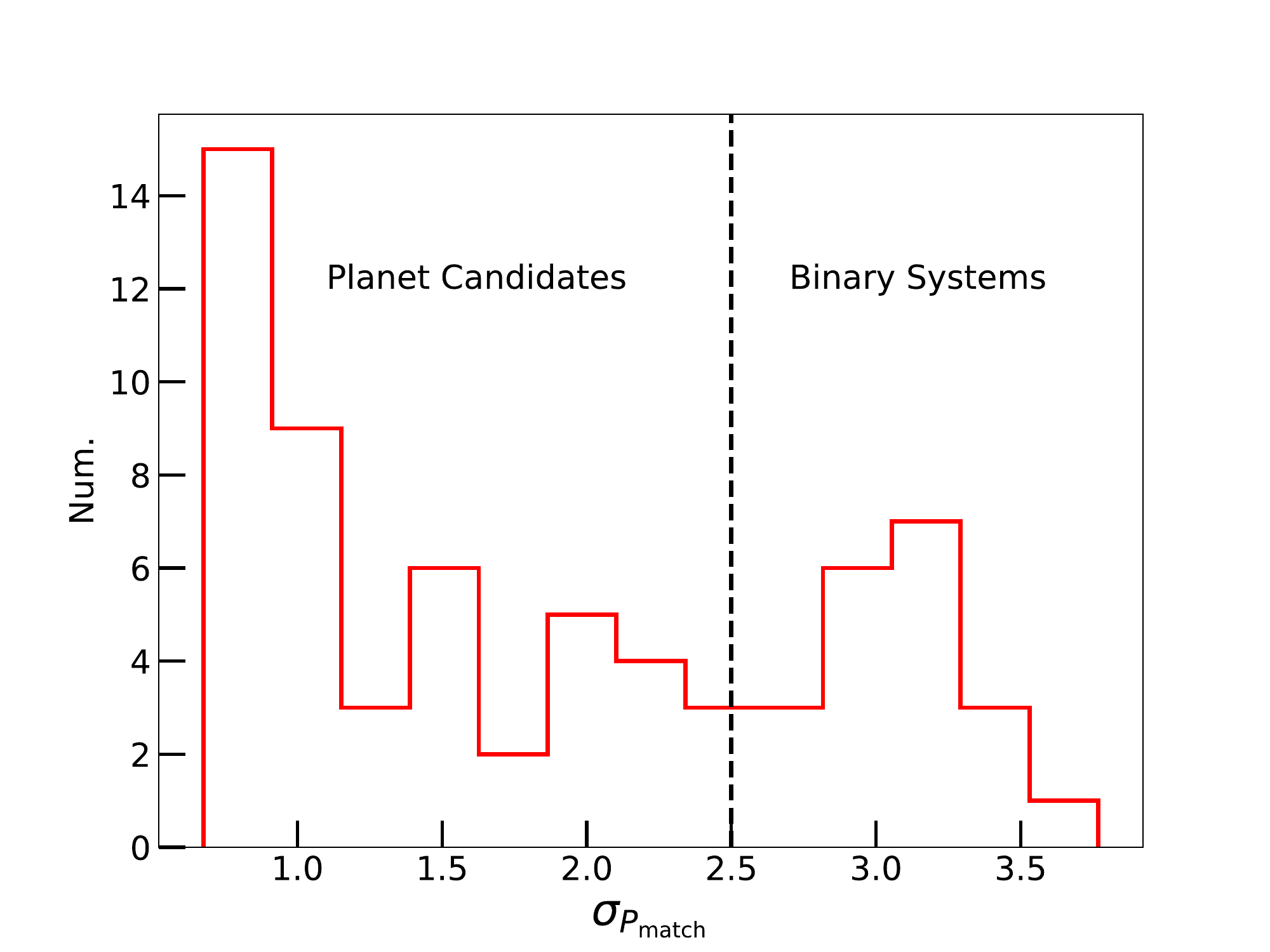}
\caption{$\sigma_{P_{\rm match}}$ distribution of 67 candidates passing the light centroid, odd/even and secondary tests as well as visual inspection. We remove candidates with $\sigma_{P_{\rm match}}\geq2.5$ as eclipsing binary systems whose stellar rotational modulation is correlated with the transit periods (or their aliases).}
\label{sigma_P}
\end{figure}

In order to remove such false positives with ellipsoidal variation, we mask out all transit signals in the raw QLP light curves of each candidate and perform a Lomb-Scargle periodogram \citep{Lomb1976,Scargle1982} analysis between 0.4 and 12 days to measure the stellar rotation period $P_{\rm rot}$. We regard the highest peak of the periodogram as the rotational period $P_{\rm rot}$. Following the methodology described in \cite{Coughlin2014}, we examine the significance of the match between $P_{\rm BLS}$ and $P_{\rm rot}$ by calculating:
\begin{equation}
    \Delta P=\frac{P_{\rm BLS}-P_{\rm rot}}{P_{\rm BLS}}
\end{equation}
and 
\begin{equation}
    \Delta P'=abs(\Delta P- int(\Delta P)),
\end{equation}
where $abs$ returns the absolute value, and $int$ yields the nearest integer. This method examines and accounts for any possible period ratios between $P_{\rm rot}$ and $P_{\rm BLS}$. We then transform $\Delta P'$ to a value quantifying the significance of the similarities between these two periods by computing the inverse of the complementary error function: 
\begin{equation}
    \sigma_{P_{\rm match}}=\sqrt{2}\times {\rm erfcinv}(\Delta P').
\end{equation}
A larger $\sigma_{P_{\rm match}}$ value means that $P_{\rm BLS}$ and $P_{\rm rot}$ are more likely to be from the same origin. Figure \ref{sigma_P} displays the $\sigma_{P_{\rm match}}$ distribution of our sample. We remove candidates with $\sigma_{P_{\rm match}}$ greater than 2.5 (roughly corresponding to a $2.5\sigma$ significance) and with the peak in the periodogram having a false alarm probability below 0.1\%. We reduce the candidate number of our sample to 44 after this step. We note that the choice of $2.5\sigma$ is somewhat arbitrary and the key point here is to exclude obvious eclipsing binary systems with ellipsoidal variations. We carry out an independent test with a $3\sigma$ threshold. With a stricter threshold, it requires a better match between the orbital and rotation period, which will exclude fewer candidates. Using a $3\sigma$ cut, we find 10 new planet candidates left in the sample. However, all these additional candidates are excluded according to the planet radius, orbital period and impact parameter cut in the final step (see Section \ref{lightcurve_modeling}). We thus consider that the choice of selection cut has little effect on the our statistics. More importantly, the false negative rate of setting a $2.5 \sigma$ cut here will be calculated and considered in the injection and recovery test (see Section \ref{injection_recovery}). 


\subsection{Ground-based Photometry}\label{photometry_check}
We next cross match these 44 candidates with the \tess\ objects of interest (TOI) catalog, and we find that 19 out of 44 candidates are known TOIs. Based on publicly available observational notes on ExoFOP\footnote{\url{https://exofop.ipac.caltech.edu/tess/}}, we further exclude two targets from our planet candidate sample. TIC 305478010 (TOI-3580) is confirmed to be a nearby eclipsing binary through the \gaia\ time-series \citep{Panahi2022}. Additionally, we also retire TIC 7439480 as the ground observations have confirmed that the signal is on the nearby star TIC 7439481 (TOI-4339). Although we utilize outside studies to reject false positives here, we emphasize that all candidates in our final catalog are vetted through ground-based photometric observations (see Section \ref{candidate_followup_observation}).


\subsection{Light Curve Modeling}\label{lightcurve_modeling}
Finally, we derive the best physical parameters of each companion. First, we apply the \code{celerite} package \citep{Foreman2017} to re-detrend the raw QLP light curve by fitting a Gaussian Process (GP) model with a simple Matern 3/2 kernel. The out-of-transit part of the total light curve is selected in the phase space using
\begin{equation}
    {\Phi} \geq \frac{\tau}{2\times P_b}+\varphi,
\end{equation}
where $\Phi$ represents the orbital phase of the planet candidate, $P_b$ and $\tau$ are the orbital period and duration. We account for the uncertainties on period $P_b$, mid-transit time $T_{0,b}$ and duration $\tau$ by including an additional factor $\varphi$, which was set to 0.02. 

After detrending, we conduct a uniform transit fit across our sample. We utilize the \code{juliet} package \citep{juliet} to perform the fit, which employs the dynamic nested sampling approach to determine the posterior distributions of each parameter based on the \code{dynesty} package \citep{Higson2019,Speagle2019}. The transit is modelled by \code{batman} \citep{Kreidberg2015}. We set Gaussian priors that center around the orbital period $P_{b}$ and mid-transit time $T_{0,b}$ found by our detection pipeline with a width of 0.2 days. We adopt a quadratic limb-darkening law for the \tess\ photometry, as parameterized by \cite{Kipping2013}, as well as an informative Gaussian prior on stellar density based on TICv8 stellar parameters \citep{Stassun2019tic}. In addition, \code{juliet} makes use of the new parametrizations $r_{1}$ and $r_{2}$ to efficiently sample points in the planet-to-star radius ratio and impact parameter space \citep{Espinoza2018}, and we place wide uniform priors on both of them. Regarding the light contamination, we set a tight truncated normal prior on the dilution factor $D_{\rm TESS}$\footnote{We convert the contamination ratio reported by TICv8 into dilution factor using Equation 6 in \cite{juliet}: $D_{\rm TESS}=1/(1+A_{\rm D})$.} with a standard deviation of 0.05, and allow it to vary between 0 and 1. We include a photometric jitter term to account for additional white noise in the \tess\ photometry and fit circular orbits with eccentricity fixed at 0. We summarize our prior settings in Table \ref{transit_fit_prior}. 

We accept candidates with orbital period $0.8 \leq P_b\leq 10$ days, impact parameter $b\leq0.9$ and companion size $7\ R_{\oplus}\leq R_{b}\leq 2\ R_{J}$, which is the parameter space of concern in this work. We end up with a final sample of nine candidates, all of which are previously alerted TOIs. Among them, four are confirmed planets, and the other five are planet candidates. Table \ref{final_candidates} lists their basic information. The other 33 targets that do not satisfy the selection limits of $R_b$, $P_b$, and $b$ are listed in the appendix Table \ref{removed_candidates_from_modeling} along with their exclusion reasons.  We show the light curves and best-fit transit models of these 33 objects in Figure \ref{removed_candidates_from_modeling_lc}.

\section{Candidate Follow-up Observations}\label{candidate_followup_observation}
We acquired ground-based time-series follow-up photometry of all of our five planet candidates as part of the \textit{TESS} Follow-up Observing Program Sub Group 1\footnote{\url{https://tess.mit.edu/followup}} \citep[TFOP SG1;][]{Collins2019} to determine the source of the signal detected in the TESS data. We used the {\tt TESS Transit Finder}, which is a customized version of the \code{Tapir} software package \citep{Jensen2013}, to schedule our transit observations. The images were calibrated and photometric data were extracted using \code{AstroImageJ} \citep{Collins2017}. We briefly summarize observations in Table \ref{po}. More details of all ground-based observations can be found in Appendix \ref{detail_ground_obs}.

The consistency of transit depth across multi-band observations reduces the chances of the blended eclipsing binary scenario within the follow-up aperture. Moreover, the transit events were all verified to occur on the target star except TIC 382602147 (i.e., TOI-2384), which has a nearby star ($\Delta T=3.64$ at $0.85\arcsec$) blended in the follow-up aperture. However, we demonstrate here that the transit happens on target. QLP reports a duration ratio of $\tau_{12}/\tau_{13}=0.364$, where $\tau_{12}$ is the ingress duration and $\tau_{13}$ represents the time span from first-to-third contact during the transit event. If the flux drop happens on the blended star, the transit depth would be limited within $\left(\tau_{12}/\tau_{13}\right)^2=0.132$ \citep{Seager2003}. Since the measured transit depth is 0.0288, the blended star would have to contribute at least 21.7\% light in the \tess\ aperture, corresponding to $T_{\rm blended}-T_{\rm target}\leq1.39$. However, the nearby star is fainter than the target with $\Delta T=3.64$, which rules out this possibility. 

Based on these ground observations, we conclude that all of the nine candidates in our vetted sample are confirmed or verified planets with a very low likelihood to be false positives, which we quantify later in the paper.

\begin{table*}
    \centering
    \caption{The nine hot Jupiter candidates around early-type M dwarfs detected by our pipeline and survived after vetting.}
    \begin{tabular}{lcccccccc}
        \hline\hline
        TIC       &TOI &Tmag   &Period (days) &Impact parameter $b$ &$R_{p}\ (R_{J})$ &$f_{{\rm Star},i}$ &$f_{\rm FP}$ &TFOP Status\\\hline
        20182780 &3984 &13.46 &$4.3534\pm0.0002$ &$0.23\pm0.10$ &$0.65\pm0.02$ &0 &0$^{[1]}$ &VPC$^{[2]}$ \\ 
        33521996 &468 &13.34 &$3.3256\pm0.0003$ &$0.46\pm0.10$ &$1.00\pm0.03$ &0 &0 &KP$^{[3]}$; \cite{Hartman2015} \\
        71268730 &5375 &12.46 &$1.7215\pm0.0001$ &$0.11\pm0.07$ &$0.90\pm0.03$ &0.030 &0.072 &VPC \\ 
        79920467 &3288 &13.30 &$1.4339\pm0.0001$ &$0.24\pm0.15$ &$0.97\pm0.03$ &0.046 &0.087 &VPC \\ 
        95057860 &4201 &13.50 &$3.5824\pm0.0003$ &$0.22\pm0.13$ &$1.05\pm0.03$ &0.056 &0.096 &VPC \\ 
        155867025 &3714 &13.18 &$2.1549\pm0.0002$ &$0.17\pm0.11$ &$1.00\pm0.03$ &0 &0 &KP; \cite{Canas2022} \\
        382602147 &2384 &13.31 &$2.1357\pm0.0001$ &$0.63\pm0.05$ &$1.09\pm0.03$ &0.064 &0.104 &VPC \\ 
        445751830 &3757 &13.19 &$3.4389\pm0.0003$ &$0.79\pm0.06$ &$1.10\pm0.03$ &0 &0 &KP; \cite{Kanodia2022} \\
        455784423 &3629 &12.79 &$3.9394\pm0.0012$ &$0.21\pm0.14$ &$0.72\pm0.02$ &0 &0 &KP; \cite{Canas2022} \\
        \hline\hline
    \end{tabular}
    \begin{tablenotes}
       \item[1]  [1]\ We set the $f_{\rm FP}$ to 0 for TOI-3984 because our NEID RV observations place a $3\sigma$ upper limit of $0.32\ M_{J}$ on the companion mass, which rules out the brown dwarf, stellar binary or triple scenario.
       \item[2]  [2]\ A verified planet candidate that passes ground-based photometric follow up observation vetting. 
       \item[3]  [3]\ A known planet. 
    \end{tablenotes}
    \label{final_candidates}
\end{table*}    

\begin{table*}
\centering
    \caption{Ground-based photometric follow-up observations for five hot Jupiter candidates.}
    \begin{tabular}{cccccccc}
        \hline\hline
        TIC &TOI &Telescope &Date (UT) &Filter  &Coverage &Observtory &Location\\\hline
        20182780 &3984 &LCOGT$^{[1]}$-1m &2022-04-14 &$i'$ &Full &LCO Teide &Spain\\
        & &OSN-1.5m &2022-05-10 &$V$ &Ingress &Sierra Nevada &Spain\\
        & &OSN-1.5m &2022-05-10 &$I$ &Ingress &Sierra Nevada &Spain\\
        & &LCOGT-1m &2022-06-06 &$g'$ &Full &LCO McDonald &USA\\
        \hline
        71268730 &5375 &GdP-0.4m &2022-03-05 &clear &Full &Grand-Pra &Switzerland\\
        & &CMO-0.6m &2022-03-31 &$R_{c}$ &Ingress &Caucasian Mountain &Russia\\\hline
        79920467 &3288 &LCOGT-0.4m &2021-06-07 &$i'$ &Full &LCO Siding Springs &Australia\\
        & &CDK20-0.5m &2021-09-02 &$\rm Lum$ &Full &El Sauce &Chile \\
        & &CDK20-0.5m &2021-10-28 &$\rm Lum$ &Full &El Sauce &Chile\\
        & &LCOGT-1m &2021-06-19 &$i'$ &Full &LCO Sutherland &South Africa\\
        & &LCOGT-1m &2022-05-16 &$g'$ &Full &LCO Cerro Tololo &Chile\\ \hline
        95057860 &4201 &LCOGT-1m &2021-09-01 &$i'$ &Ingress &LCO Siding Springs &Australia\\
        & &LCOGT-1m &2021-09-26 &$g'$ &Ingress &LCO Sutherland &South Africa\\
        & &LCOGT-1m &2021-09-26 &$i'$ &Ingress &LCO Sutherland &South Africa\\
        & &LCOGT-1m &2021-10-13 &$g'$ &Full &LCO Cerro Tololo &Chile\\
        & &LCOGT-1m &2021-10-13 &$i'$ &Full &LCO Cerro Tololo &Chile\\\hline
        382602147 &2384 &CDK14-0.36m &2020-11-09 &$R_{c}$ &Full &El Sauce &Chile\\
        & &LCOGT-1m &2021-08-05 &$g'$ &Full &LCO Cerro Tololo &Chile\\
        \hline\hline
    \end{tabular}
    \begin{tablenotes}
     \item[1]  [1]\ Las Cumbres Observatory Global Telescope \citep[LCOGT;][]{Brown2013}. 

    \end{tablenotes}
    \label{po}
\end{table*}

\section{Injection and Recovery}\label{injection_recovery}

In this section, we measure the sensitivity of our detection pipeline and quantify the completeness of our final planet candidate sample through injection and recovery tests. We insert planet signals into the spline model detrended light curves (see Section \ref{light_curve_preprocessing}) and feed these synthetic data to our planet detection pipeline. Since the detrended light curves have already passed the low resolution BLS search, which resulted in non-detections, the newly alerted events in this experiment would be the signals we injected rather than unexpected detrending issues as we mentioned in Section \ref{vetting}. 

\subsection{Sensitivity of the Detection Pipeline}\label{injection}
The injection is carried out as followed:
\begin{enumerate}[(i)]
\item We uniformly divide the period-radius space ($0.8 \leq P_b\leq 10$ days, $7\ R_{\oplus}\leq R_{p}\leq 2\ R_{J}$) into a $5\times5$ grid;
\item Within each cell, we draw 20 sets of physical parameters $P_b$, $R_{p}$ as well as impact parameter $b\leq 0.9$ from uniform distributions, and randomly generate mid-transit times $T_{0,b}$ between the start time of a light curve $t_{\rm begin}$ and $t_{\rm begin}+P_b$;
\item We build artificial transit models using \code{batman} assuming circular orbits, during which we correct the dilution effect for each star\footnote{The planet-to-star radius ratio we set is $(R_{p}/R_{\ast})\times D_{\rm TESS}^{0.5}$, where $R_p$ and $R_{\ast}$ are the injected planet size and stellar radius, $D_{\rm TESS}$ is the dilution factor.}. We fix the limb-darkening coefficients [$\mu_1,\mu_2$] to [0.3, 0.3] in this step for simplicity;
\item We initialize the model at a high cadence level (100,000 points) and resample it to the real observation time stamps, and superimpose the transit model with the detrended light curve. 
\end{enumerate}


We randomly choose 3,000 stars from the 60,382 stars in our parent sample without transit alerts and apply the above injection process. We put all synthetic light curves through our detection pipeline, and record signals that are recovered. Since there are inevitably variable stars in the randomly selected 3000 stars, we also require all recovered planets to pass the ``synchronization'' test as we did in Section \ref{vetting}. We did not perform the odd/even and secondary eclipse analysis because 1) the false negative rate of this analysis in the vetting step is negligible (see Section \ref{secondary_odd_even}); 2) here we only inject periodic signals with consistent transit depth. 

\begin{figure}
\includegraphics[width=0.49\textwidth]{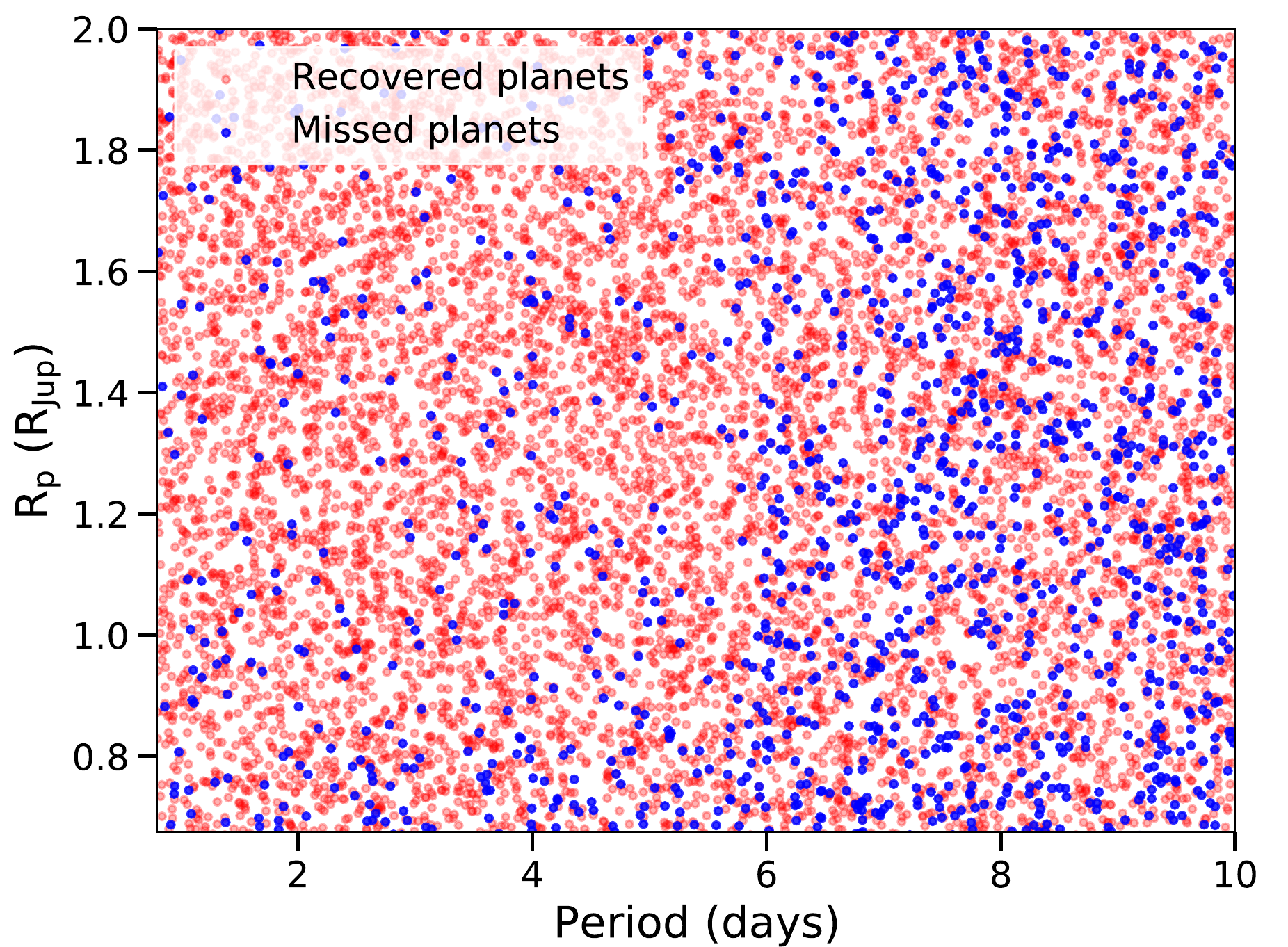}
\caption{Period and planet radius of 10,000 randomly selected injected planets. Red dots mean the recovered planets while blue dots represent the missed planets during the injection and recovery experiment.}
\label{sampled_planets}
\end{figure}

\begin{figure*}
\includegraphics[width=0.495\textwidth]{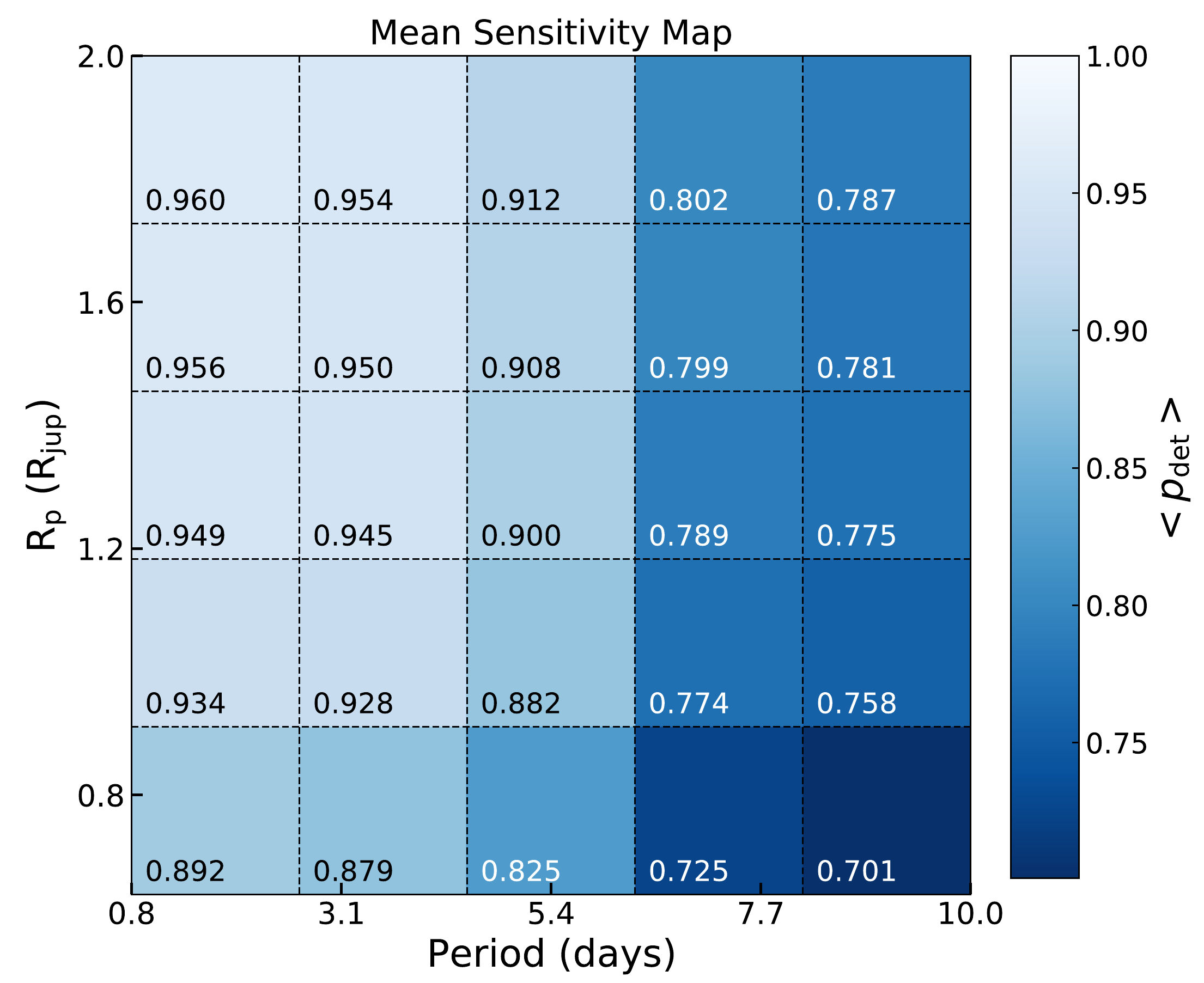}
\includegraphics[width=0.489\textwidth]{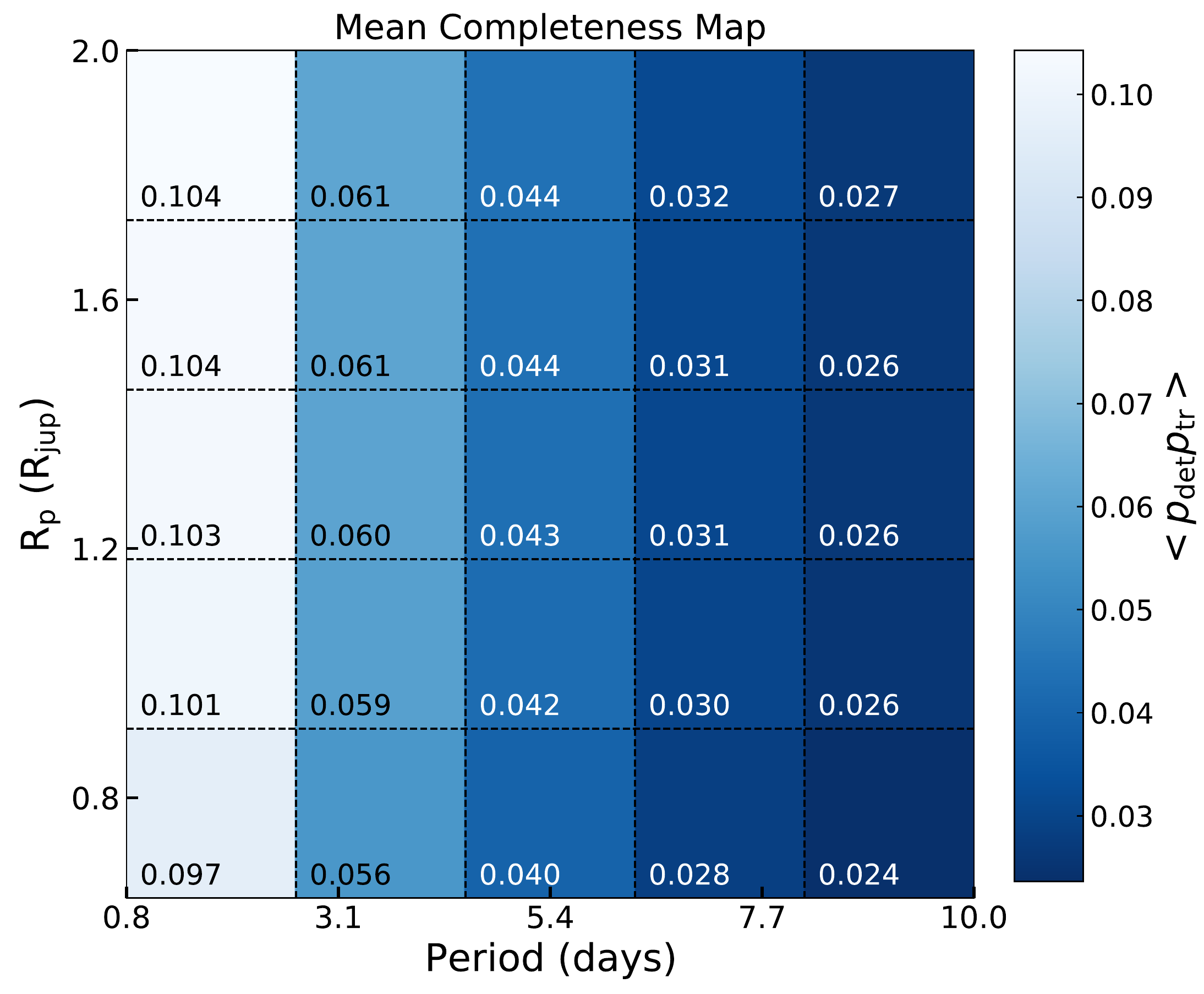}
\caption{{\it Left panel}: The average sensitivity ($\left<p_{\rm det}\right>$) of our detection pipeline as a function of orbital period and planet radius based on 3,000 stars randomly drawn from the full sample (60,819 stars), each with 500 injected signals randomly distributed in the period-radius space. {\it Right panel}:  The average search completeness ($\left<p_{\rm det}p_{\rm tr}\right>$) map of the same sample after accounting for both the pipeline sensitivity and the transit probability. Lighter colors indicate higher detection sensitivity or higher completeness, with the numerical values labeled within each cell.}
\label{detection_completeness_map}
\end{figure*}

Consequently, we insert and test 1,500,000 signals in total. We show the distribution of a total of 10,000 recovered or missed planets randomly drawn from this simulation as a function of planet period and radius in Figure \ref{sampled_planets}. Based on the fraction of recovered planets in each cell, we generate the individual sensitivity maps ($p_{{\rm det},i}$) for these 3,000 random stars, and combine all of them to provide an average transit detection sensitivity ($\left<p_{\rm det}\right>$) map as in the left panel of Figure \ref{detection_completeness_map}. We also conduct a study on injecting the planet signals into the raw QLP light curves instead of the detrended data sets, followed by spline model detrending and planet searching. We find a minor average BLS SNR decrease of $\lesssim 1$. The difference in the final mean sensitivity map is around 0.003, which is within the errors we consider below (see Sections \ref{completness_correction} and \ref{occurrence_rate_cell}). 

\subsection{Completeness of the Planet Candidate Sample}\label{completness_correction}

Before deriving the planet occurrence rate, we have to correct for the geometric probability of transit for the detectability map to find out our sample completeness. Based on Kepler's Third Law, the transit probability is defined as 
\begin{equation}
    p_{\rm tr}=0.9 \frac{R_{\ast}}{a}=0.9 R_{\ast}  \left( \frac{GM_{\ast}P^{2}}{4 \pi^{2}}\right)^{-1/3},
\end{equation}
where $R_\ast$ and $M_\ast$ are the radius and mass of the star, $P$ is the orbital period of the companion in a circular orbit. We include a factor of 0.9 since we only take planet candidates with $b\leq 0.9$ into consideration in this study. For each injected planet of every randomly selected star in Section \ref{injection}, we compute the transit probability and multiply this factor in the detectability map to account for the geometric effect. We generate individual completeness ($p_{{\rm det},i}p_{{\rm tr},i}$) map for each star, and show the resulting average ($\left<p_{\rm det}p_{\rm tr}\right>$) map in the right panel of Figure \ref{detection_completeness_map}. We rerun the injection and recovery process using another two different sets of 3,000 stars, and find that the differences in the completeness map are all within $0.004$. We account for this uncertainty on $\left<p_{\rm det}p_{\rm tr}\right>$ in the occurrence rate computation. 

\section{Occurrence Rate}\label{occurrence_rate_cell}

\begin{figure*}
\includegraphics[width=\textwidth]{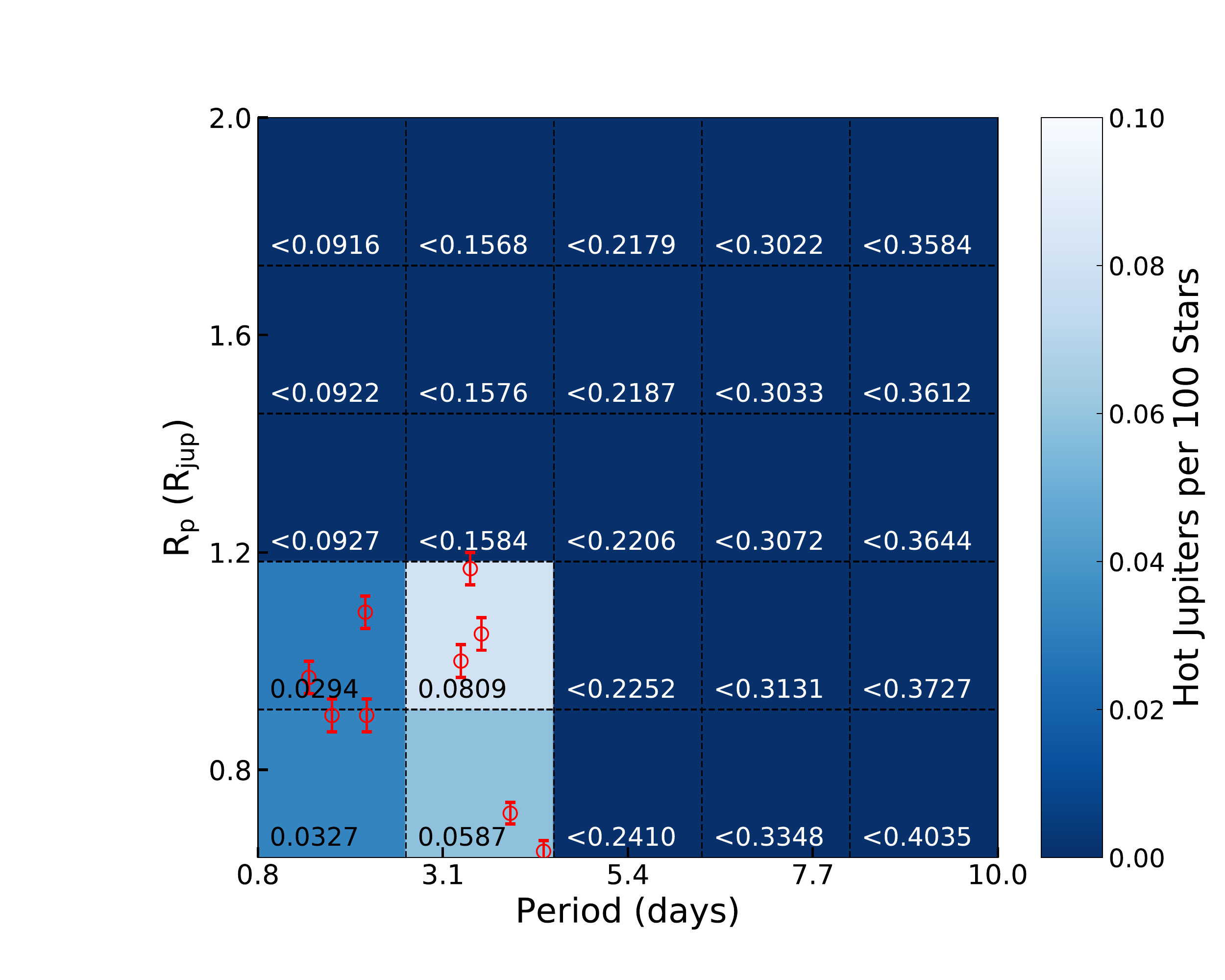}
\caption{The planet occurrence rate (number of planets per 100 stars) as a function of orbital period and planet radius. Red circles are the nine planet candidates identified by our pipeline after vetting. The occurrence rates or the $3\sigma$ upper limits are labeled in each cell (see Section \ref{occurrence_rate_cell}).}
\label{occurrence_map}
\end{figure*}

We measure the occurrence rate by counting the number of observed planets within each cell in the $5\times5$ period-radius grid and dividing it by the summed completeness map. The summed completeness map is constructed through multiplying the average completeness map with the total star number (60819). Due to the small errors on the period and the companion radius as well as the relatively large grid size, we ignore the uncertainties from $P_{b}$ and $R_{b}$, and we only consider Poisson errors from counting and the uncertainties from $\left<p_{\rm det}p_{\rm tr}\right>$. Finally, except for the best estimate for the occurrence rate, we also provide the upper and lower bounds by considering extreme cases when candidates are all real planets or false positives. 

We define an effective number of stars $n_{\rm trial}$ after correcting the sample completeness following \cite{Petigura2018} and \cite{Zhou2019} as 
\begin{equation}
    n_{\rm trial}=n_{\star}\left<p_{\rm det}p_{\rm tr}\right>,
\end{equation}
where $n_{\star}$ is the total star number used in this study (60,819).
In addition, we calculate the number of observed planets $n_{\rm obs}$ as
\begin{equation}
    n_{\rm obs}=\Sigma_{i=1}^{n_{p}}(1-f_{{\rm FP},i}),
\end{equation}
where $f_{{\rm FP},i}$ is the false positive rate of each planet candidate found by our detection pipeline, $n_{p}$ is the number of total candidates. We set $f_{{\rm FP},i}$ to zero for four confirmed planets as well as TIC 20182780 (TOI-3984), which was one of the five candidates but validated by our NEID spectroscopic observations (see Appendix \ref{TOI3984}). For the other four verified Jupiter candidates, though ground-based photometry has confirmed the signal on target, they still have a chance to be low mass M-type stars or brown dwarfs, all of which have similar size. We obtain this $f_{{\rm FP},i}$ factor through two steps. First, we utilize the \code{Forecaster} \citep{Chen2017} package to estimate the probability of each candidate $i$ that being a star with mass above $0.08\ M_{\odot}$ given a measured radius ($f_{{\rm Star},i}\sim 5\%$). Next, we estimate the probability ($f_{\rm BD}$) of the companion being a brown dwarf ($13.6\ M_{J}\leq M_{p}\leq 80\ M_{J}$) instead of a real planet ($M_{p}<13.6\ M_{J}$). To do this, we retrieve all objects with radius $7\ R_{\oplus}\leq R_{p}\leq 2\ R_{J}$ that have been detected so far, and we find 511 hot Jupiters and 23 brown dwarfs. Thus, we derive a brown dwarf probability of 4.3\%. The final $f_{{\rm FP},i}$ factor is set based on the results from the two steps above:
\begin{equation}
    f_{{\rm FP},i}=f_{{\rm Star},i}+(1-f_{{\rm Star},i})\times f_{\rm BD}.
\end{equation}
The occurrence rate in a cell with real planet candidate detections can thus be computed as 
\begin{equation}
    f_{\rm cell}=n_{\rm obs}/n_{\rm trial}.\label{eqn:occ-rate}
\end{equation}

Assuming the occurrence rate of each cell is $f_{\rm cell}$, the probability to detect $d\ (d\leq n_{\rm obs})$ planets in a specific cell follows a binomial distribution \citep{Burgasser2003,Petigura2018}:
\begin{equation}
    P(n_{\rm trial}, d, f_{\rm cell})=Nf_{\rm cell}^{d}\left(1-f_{\rm cell}\right)^{n_{\rm trial}-d},
\end{equation}
where
\begin{equation}
    N=\frac{\Gamma(n_{\rm trial}+1)}{\Gamma(d+1)\Gamma(n_{\rm trial}-d+1)}.
\end{equation}
If there is a null detection in a cell, we estimate a $3\sigma$ upper limit on the occurrence rate through
\begin{equation}
    \int_{0}^{f_{\rm cell,max}} (n_{\rm trial}+1)P(n_{\rm trial}, 0, f_{\rm cell})\ df_{\rm cell}=C,
\end{equation}
where $C$ is the confidence interval (99.7\%).
Therefore, the maximum occurrence rate $f_{\rm cell,max}$ in a cell with non-detection can be analytically solved as 
\begin{equation}
    f_{\rm cell,max}=1-(1-C)^{1/(n_{\rm trial}+1)}.
\end{equation}

Figure \ref{occurrence_map} shows the cell-by-cell planet occurrence rate. Based on the results of each cell, we next calculate an average completeness value over the $5\times5$ grid. We run Monte Carlo simulations to estimate $\sigma_{n_{\rm obs}}$. Overall, we determine a total average occurrence rate of $0.27\pm0.09\%$, where the error mainly comes from the Poisson uncertainty. Since we are unclear about the nature of five planet candidates (including TOI-3984 as we did not measure the orbit), we also estimate the upper and lower limits of the occurrence rate by assuming all candidates are true planets and false positives. This way, we obtain a conservative upper bound of $0.29\%$ and a conservative lower bound of $0.13\%$.

\section{Discussion}\label{discussion}

\subsection{Comparison to hot Jupiters around AFGK dwarfs}

\begin{figure}
\includegraphics[width=0.49\textwidth]{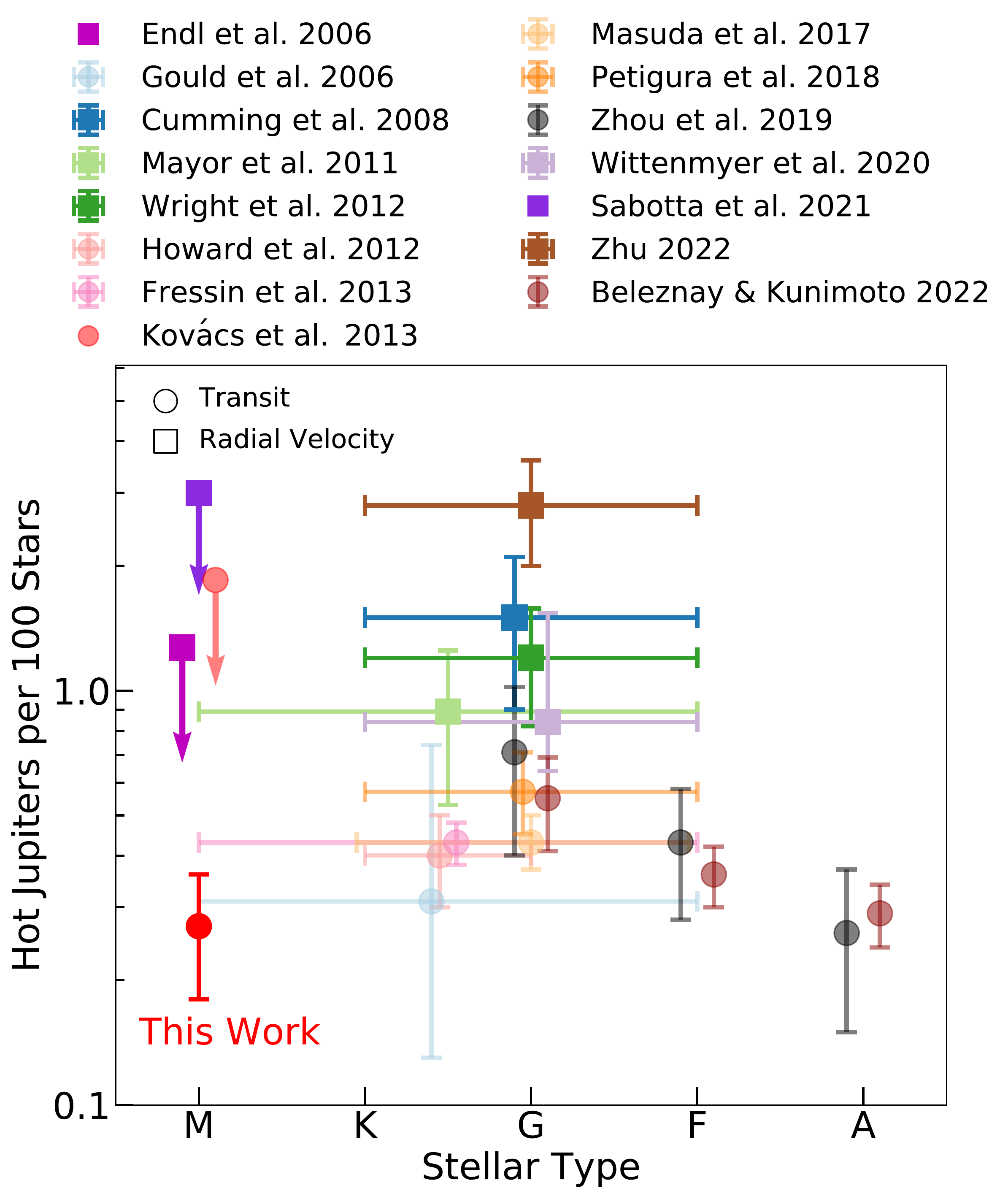}
\caption{Occurrence rates of hot Jupiters as a function of stellar type. Different colors represent different studies. Results from transit and RV surveys are shown as circles and squares, respectively. The horizontal uncertainties mark the range of stellar type used in each work. We added small random shifts to the horizontal coordinates for the studies on FGKM stars for clarity.}
\label{occ_stellar_mass}
\end{figure}

Compared with the occurrence rates of hot Jupiters around AFGK stars, we find the value $0.27\pm0.09\%$ for early-type M dwarfs deviates from the majority of measurements. We note that our result is within the occurrence rate upper limits of Jovian-size planets around M dwarfs previously reported by \cite{Endl2006}, \cite{Kovacs2013} and \cite{Sabotta2021}. Figure \ref{occ_stellar_mass} shows the hot Jupiter occurrence rates from different works as a function of stellar type. We caution the readers that these studies use different methods (transit or RV) and have slightly different definitions for hot Jupiters. A summary of these results is presented in Table \ref{occ_results_other_works}. After adding a measurement at the low stellar mass end from this work, the occurrence rate of hot Jupiters appears to have a maximum peak around G stars and decrease towards M and A dwarfs, but actually most measurements still agree with each other within 1--2$\sigma$ so we cannot draw definitive conclusions regarding the trend in the hot Jupiter occurrence across stellar types. 

However, if this occurrence rate trend is real, it might reflect the different formation history of hot Jupiters around different types of stars. Since the mass of the protoplanetary disk scales linearly with the stellar mass $M_{\ast}$ \citep{Andrews2013}, theoretical works predict that Jupiters are more rare around M dwarfs \citep{Laughlin2004,Ida2005,Kennedy2008,Liu2019} due to the shortage of solid materials in the protoplanetary disks to support giant planet formation. Indeed, a simulation carried out by \cite{Burn2021} shows that gas giants ($M_{p}>100\ M_{\oplus}$) cannot form around M dwarfs with $M_{\ast}<0.5\ M_{\odot}$ through core accretion. Such drawback could, in principle, be compensated by metal-rich stars \citep{Maldonado2020}. For A-type stars, if there is indeed a drop in the occurrence rate of hot Jupiters, it could be attributed to several potential reasons. First, the rapid rotation of A-type stars and their high surface temperatures would impede the giant planet detection and confirmation through spectroscopic observations. In addition, hot Jupiters may be engulfed by their host A stars \citep{Stephan2018}. Finally, the disk lifetime of A stars tends to be shorter than that of FGK stars \citep{Ribas2015} and there could be fewer successfully formed giant planets before disk dissipation. More detections and studies on hot Jupiters around A stars are required to draw conclusions.


As can be seen from Figure \ref{occ_stellar_mass}, the occurrence rates reported by RV surveys, although consistent within $1\sigma$, are systematically higher than the values from transit studies (see \citealt{Zhu2021} and references therein). In particular, recent work by \cite{Zhu2022} used the Sun-like sample from the California Legacy Survey \citep[CLS;][]{Rosenthal2021} and measured a hot Jupiter frequency of $2.8\pm0.8\%$, which is substantially higher than the rate obtained in our work around early-type M dwarfs. \cite{Wright2012} pointed out that such a difference between the RV and transit results might be partly owing to different stellar metallicity between these two samples. However, a further study of the \kepler\ stellar sample from \cite{Dong2014} shows that they have a sub-solar metallicity ($\sim-0.04$ dex) similar to the RV sample ($\sim0.0$ dex), which implies that metallicity may have minor impact on this discrepancy. A similar conclusion was also drawn by \cite{Guo2017}. Moreover, according to the statistics from \cite{Moe2021}, RV surveys probably increase the detection rates of hot Jupiters by a factor of $1.8\pm0.2$ by removing spectroscopic binaries among their parent samples, which could result in this feature. A promising way to test this hypothesis is to search for close stellar companions of transiting hot Jupiters with high contrast imaging \citep[e.g.,][]{Ngo2016}, excluding circumbinary systems, and compare the remaining sample with the RV sample. Finally, stellar age may also play a role although such an effect has not been thoroughly discussed \citep{Donati2016}. 

\begin{table*}
    \centering
    \caption{A summary of occurrence rates of hot Jupiters from other works}
    \begin{tabular}{lcccc}
        \hline\hline
        Work       &$f_{\rm occ} (\%)$ &Stellar type &Method   &Definition of hot Jupiters\\\hline
        \cite{Endl2006} &$<1.27\%$ &M &RV &$M_{p}\sin i\sim 1\ M_{J}$, $\rm a<1\ AU$\\
        \cite{Gould2006} &$0.31^{+0.43}_{-0.18}$ &FGKM &Transit &$1\leq R_{p}\leq 1.25\ R_{J}$, $3\leq P_{b}\leq5 $ days\\
        \cite{Cumming2008} &$1.5\pm0.6$ &FGK &RV &$M_{p}\sin i>0.3\ M_{J}$, $P_{b}<11.5$ days\\
        \cite{Mayor2011} &$0.89\pm0.36$ &FGKM &RV &$M_{p}\sin i>50\ M_{\oplus}$, $P_{b}<11$ days\\
        \cite{Wright2012} &$1.20\pm0.38$ &FGK &RV &$M_{p}\sin i>0.1\ M_{J}$, $P_{b}<10$ days\\
        \cite{Howard2012} &$0.4\pm0.1$ &GK &Transit &$8\leq R_{p}\leq 32\ R_{\oplus}$, $P_{b}<10$ days\\
        \cite{Fressin2013} &$0.43\pm0.05$ &FGKM &Transit &$6\leq R_{p}\leq 22\ R_{\oplus}$, $0.8\leq P_{b}\leq 10$ days\\
        \cite{Kovacs2013} &$<1.7$-$2.0\%$ &M &Transit &$R_{p}\sim 1.0\ R_{J}$, $0.8\leq P_{b}\leq10$ days\\
        \cite{Masuda2017} &$0.43^{+0.07}_{-0.06}$ &FGK &Transit &$0.8\leq R_{p}\leq 2\ R_{J}$, $P_{b}< 10$ days\\
        \cite{Petigura2018} &$0.57^{+0.14}_{-0.12}$ &FGK &Transit &$8\leq R_{p}\leq 24\ R_{\oplus}$, $1\leq P_{b}\leq 10$ days\\
        \cite{Zhou2019}$^{[1]}$ &$0.26\pm0.11$ &A &Transit &$0.8\leq R_{p}\leq 2.5\ R_{J}$, $0.9\leq P_{b}\leq 10$ days\\
        \cite{Zhou2019} &$0.43\pm0.15$ &F &Transit &$0.8\leq R_{p}\leq 2.5\ R_{J}$, $0.9\leq P_{b}\leq 10$ days\\
        \cite{Zhou2019} &$0.71\pm0.31$ &G &Transit &$0.8\leq R_{p}\leq 2.5\ R_{J}$, $0.9\leq P_{b}\leq 10$ days\\
        \cite{Wittenmyer2020} &$0.84^{+0.70}_{-0.20}$ &FGK &RV &$M_{p}\sin i>0.3\ M_{J}$, $1 \leq P_{b}\leq 10$ days\\
        \cite{Sabotta2021} &$<3\%$ &M &RV &$100<M_{p}<1000\ M_{\oplus}$, $P_{b}<10$ days\\
        \cite{Zhu2022} &$2.8\pm0.8$ &FGK &RV &$M_{p}\sin i>0.3\ M_{J}$, $a\leq0.1$ AU\\
        \cite{Beleznay2022}$^{[2]}$ &$0.29\pm0.05$ &A &Transit &$0.8\leq R_{p}\leq 2.5\ R_{J}$, $0.9\leq P_{b}\leq 10$ days\\
        \cite{Beleznay2022} &$0.36\pm0.06$ &F &Transit &$0.8\leq R_{p}\leq 2.5\ R_{J}$, $0.9\leq P_{b}\leq 10$ days\\
        \cite{Beleznay2022} &$0.55\pm0.14$ &G &Transit &$0.8\leq R_{p}\leq 2.5\ R_{J}$, $0.9\leq P_{b}\leq 10$ days\\
        This work &$0.27\pm0.09$ &M &Transit &$
        7\ R_{\oplus}\leq R_{p}\leq 2\ R_{J}$, $0.8 \leq P_{b}\leq 10$ days\\
        \hline\hline
    \end{tabular}
    \begin{tablenotes}
      \item[1]  [1]\ \cite{Zhou2019} also reported an average occurrence rate of $0.41\pm0.10\%$ within their full AFG sample.
      \item[2]  [2]\ \cite{Beleznay2022} also reported an average occurrence rate of $0.33\pm0.04\%$ within their full AFG sample.
    \end{tablenotes}
    \label{occ_results_other_works}
\end{table*}   

\subsection{Comparison to cold Jupiters around M dwarfs}
\begin{figure}
\includegraphics[width=0.49\textwidth]{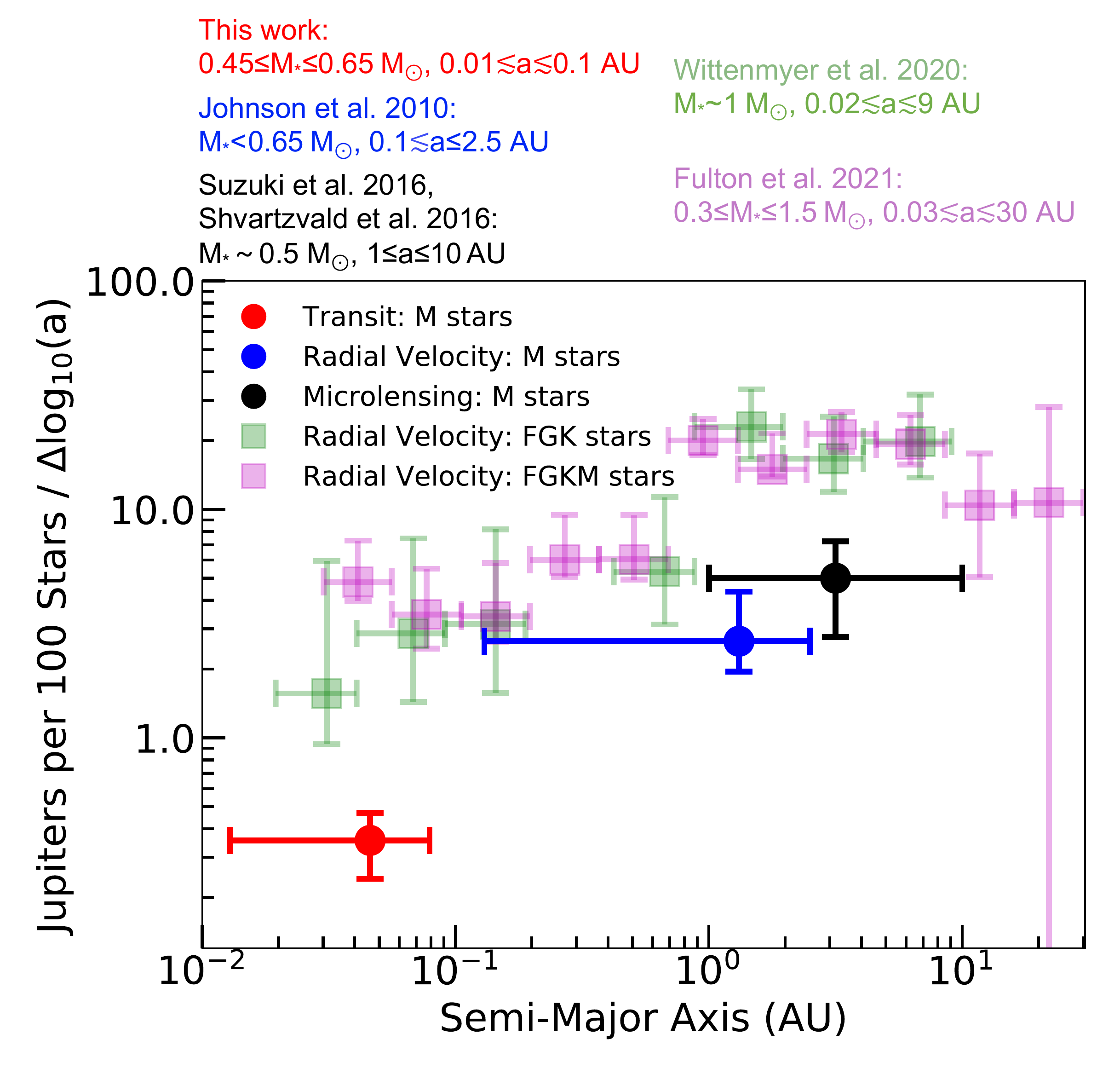}
\caption{The occurrence rate of Jupiters around M dwarfs per logarithmic semi-major axis bin (${{\rm d}N}/{\rm d}\log_{10} a$) as a function of semi-major axis. Different colors represent results from different works. The horizontal uncertainties mark the range of semi-major axis of the planet sample in each study. The reference, host star mass and semi-major axis ranges are labeled on the top of this plot. For comparison, the occurrence rates of Jupiters around FGK and FGKM dwarfs studied by \citet{Wittenmyer2020} and \citet{Fulton2021} from RV surveys are shown as green and magenta translucent squares.}
\label{occ_a}
\end{figure}

Previous research found that cold Jupiters around M dwarfs with semi-major axis $a\gtrsim 1$ AU have an occurrence rate of $\sim 4\%$. Long-term RV observations from the California Planet Survey showed that the frequency is around $3.4^{+2.2}_{-0.9}\%$ for an M dwarf ($M_{\ast}<0.6\ M_{\odot}$) harboring planets with $M_{p}>0.3\ M_{J}$ within 2.5 AU \citep{Johnson2010}. Although \cite{Johnson2010} did not claim the inner bound of their detection limit, the planet GJ 876 c in their sample has the smallest semi-major axis, about 0.13 AU \citep{Marcy2001,Rivera2005}. Furthermore, various microlensing studies reported a value around $5\%$ at 1--10 AU \citep{mufun,Cassan2012,Suzuki2016,Wise}. Although RV surveys mainly focus on giant planets within 2.5 AU while the microlensing method is sensitive to planets beyond the snow line, the occurrence rates of cold Jupiters measured using these two methods are consistent with each other at $1\sigma$. For hot Jupiters located at a distance of $a\lesssim 0.1$ AU from their early-type M dwarf hosts, we measure an occurrence rate of $0.27\pm0.09\%$. Our result is significantly smaller than the frequency of outer cold gas giants, indicating that cold Jupiters are more common than hot Jupiters around M dwarfs, which is consistent with solar-like stars \citep[e.g.,][]{Wittenmyer2020}. We show the occurrence rates per semi-major axis bin obtained using different methods as a function of the semi-major axis in Figure \ref{occ_a}. 

Combining archival data from the Anglo-Australian Planet Search \citep{Tinney2001}, \cite{Wittenmyer2020} investigated the occurrence rate of giant planets ($M_{p}>0.3\ M_{J}$) around solar-like stars across a wide range of semi-major axis ($0.02 \lesssim a\lesssim 9$ AU). More recently, \cite{Fulton2021} also looked into the same problem using an independent sample from the California Legacy Survey, of which the orbital separation spans 0.03--30 AU. The results from both works infer that the occurrence rate decreases by about 6 times from cold ($1 \lesssim a\lesssim 10$ AU, $\frac{{\rm d}N}{{\rm d}\log_{10}(a)}\sim 19\%$) to hot Jupiters ($0.01\lesssim a\lesssim 0.1$ AU, $\frac{{\rm d}N}{{\rm d}\log_{10}(a)}\sim 3\%$). In contrast, for equivalent systems around early-type M dwarf hosts, we find a steeper decrease, of about 14 times, for Jupiters at $1 \lesssim a\lesssim 10$ AU and $0.01 \lesssim a\lesssim 0.1$ AU, from $\frac{{\rm d}N}{{\rm d}\log_{10}(a)}\sim 5\%$ to $\frac{{\rm d}N}{{\rm d}\log_{10}(a)}\sim 0.34\%$ (see Figure \ref{occ_a}). The decrease we find hints that hot Jupiters around M dwarfs may be even more difficult to form than cold Jupiters when compared with G dwarfs. However, due to large uncertainties on the occurrence rates of cold Jupiters and especially that the measurements of M dwarfs come from different methods that might have sample biases, we cannot draw firm conclusions yet. Future homogeneous near-infrared spectroscopic surveys (e.g., \citealt{Mahadevan2014,Fouque2018,Reiners2018}), which perform long-term RV observations,  will shed some light on this puzzle.

Both \cite{Wittenmyer2020} and \cite{Fulton2021} reported that there exists an occurrence rate transition point around 1 AU for giant planet around FGK stars (see Figure \ref{occ_a}). This jump is suggested to be relevant to the location of the snow line \citep{Ida2008}, as an enhanced solid density beyond the snow line will facilitate the formation of the solid core under the core accretion paradigm. Due to lower irradiation, the snow line of M dwarfs is closer to the star compared with FGK dwarfs. Combining the radial velocity and microlensing findings, we can see that the occurrence rate trend of giant planets around M dwarfs is a monotonic increase as a function of semi-major axis, similar to that of the FGK dwarfs. If a sudden increase in the occurrence rate of giant planets around M dwarfs indeed exists, the exact position of this transition is still unclear due to the limited amount of data at the moment. If future occurrence rate studies on the warm Jupiters around M dwarfs confirm the existence of a transition, it would indicate that the formation of giant planets around M dwarfs is similar to FGK stars.


\subsection{New Planet Candidates}

During the candidate search, we found 7 new planet candidates that were not alerted as TOIs previously. They pass all vetting steps including centroid analysis, visual inspection for odd/even and secondary signals as well as synchronization test. All of these candidates have a radius below $7\ R_{\oplus}$ so they are not included in the statistical sample. We summarize their properties in Table \ref{new_candidates}. Since we only report the results from our uniform fits (see Section \ref{vetting}), we emphasize that more detailed analyses are required to evaluate the robustness of these candidates but such works are beyond the scope of this study. Note that during the writing of this manuscript, TIC 291109653 was alerted by the QLP faint star search program \citep{Kunimoto2022} through a Sector-combined analysis (Sectors 23 and 46), designated as TOI-5486. In this work, we independently find the signal using Sector 23 data. 


\begin{table*}
    \centering
    \caption{New planet candidates detected by our pipeline that were not announced as TOIs before. The physical parameters come from our uniform transit fit.}
    \begin{tabular}{lcccccc}
        \hline\hline
        TIC      &Tmag   &Period (days) &T0 (BJD-2457000) &$R_{p}\ (R_{J})$ &Centroid Shift (pixels) &$\sigma_{P_{\rm match}}$ \\\hline
        32296259 &12.28 &2.77757 &1492.9043  &0.42 &0.19 &0.99 \\ 
        101736867 &13.09 &2.64795 &1655.0761 &0.59 &0.83 &0.72 \\ 
        115524526 &12.94 &4.65712 &1956.4309 &0.45 &0.61 &1.99  \\ 
        246974219 &12.29 &1.90943 &1793.5922 &0.42 &0.48 &0.96  \\ 
        291109653$^{[1]}$ &12.29 &2.02479 &1929.7720 &0.33 &0.17 &0.82 \\ 
        367411575 &13.25 &1.19342 &1792.7369 &0.57 &0.48 &1.91  \\
        371315491 &13.26 &0.40622 &1571.8947 &0.55 &0.78 &1.26  \\
        \hline\hline
    \end{tabular}
    \begin{tablenotes}
       \item[1]  [1]\ TIC 291109653 was recently alerted as TOI-5486 by the QLP faint star search program \citep{Kunimoto2022}. 
    \end{tablenotes}
    \label{new_candidates}
\end{table*}    

\section{Conclusion}\label{conclusion}

In this work, we measure the occurrence rate of hot Jupiters around early-type M dwarfs as a function of orbital period and planet radius based on the observations from the \tess\ Primary Mission. Our detection pipeline includes the BLS algorithm to blindly search for giant planets among a magnitude-limited M dwarf sample of 60,319 stars. We find a total of 437 possible candidates. After investigating the centroid shifts, odd/even and secondary eclipse signals, stellar rotation and follow-up data, we identify 9 hot Jupiter candidates that are located within our planet radius and orbital period parameter space. All of them were previously announced as TOIs. We characterize the completeness of our detection pipeline through injection and recovery tests. We obtain an average occurrence rate of $0.27\pm0.09$ hot Jupiter with period $0.8 \leq P_{b}\leq 10$ days and radius $7\ R_{\oplus}\leq R_{p}\leq 2\ R_{J}$ per 100 early-type M dwarfs ($0.45 \leq M_{\ast}\leq 0.65\ M_{\odot}$). Compared with previous studies, our occurrence measurement is smaller than all measurements for FGK stars but consistent within 1--2$\sigma$. We tentatively find that the occurrence rate of hot Jupiters has a peak at G dwarfs and falls towards both hotter and cooler stars. Combining results from transit, radial velocity and microlensing surveys, we find a hint that hot Jupiters seem to struggle even more to form around M dwarfs in comparison with FGK stars. There is a possible steeper decrease for the occurrence rate per logarithmic semi-major axis bin (${{\rm d}N}/{\rm d}\log_{10} a$) of Jupiters around M dwarfs from $1\lesssim a\lesssim 10$ AU to $0.01\lesssim a\lesssim 0.1$ AU in contrast to FGK stars. We also report seven new planet candidates with planet radius below $7\ R_{\oplus}$ that were newly identified in this work, which require detailed analysis and further follow-up data to confirm their planetary nature. 

\section{Acknowledgments}
We thank the anonymous referees for their comments that improved the quality of this publication. We thank Johanna~K.~Teske for reviewing this paper and providing corrections. We are grateful to Weicheng Zang for helpful discussions regarding the statistical results from microlensing, and Michelle Kunimoto for useful discussions on the QLP faint star search program.

This work is partly supported by the National Science Foundation of China (Grant No. 12133005).
This research uses data obtained through the Telescope Access Program (TAP), which has been funded by the TAP member institutes.
The authors acknowledge the Tsinghua Astrophysics High-Performance Computing platform at Tsinghua University for providing computational and data storage resources that have contributed to the research results reported within this paper.
A.A.B. and N.A.M acknowledge the support of the Ministry of Science and Higher Education of the Russian Federation under the grant 075-15-2020-780 (N13.1902.21.0039).
This work makes use of observations from the LCOGT network. Part of the LCOGT telescope time was granted by NOIRLab through the Mid-Scale Innovations Program (MSIP). MSIP is funded by NSF.
We acknowledge the use of \tess\ public data from pipelines at the \tess\ Science Office and at the \tess\ Science Processing Operations Center. 
We acknowledge the use of TESS High Level Science Products (HLSP) produced by the Quick-Look Pipeline (QLP) at the TESS Science Office at MIT, which are publicly available from the Mikulski Archive for Space Telescopes (MAST). Funding for the TESS mission is provided by NASA's Science Mission directorate.
Resources supporting this work were provided by the NASA High-End Computing (HEC) Program through the NASA Advanced Supercomputing (NAS) Division at Ames Research Center for the production of the SPOC data products.
This research has made use of the Exoplanet Follow-up Observation Program website, which is operated by the California Institute of Technology, under contract with the National Aeronautics and Space Administration under the Exoplanet Exploration Program. 
This paper includes data collected by the \tess\ mission, which are publicly available from the Mikulski Archive for Space Telescopes\ (MAST).
This work has made use of data from the European Space Agency (ESA) mission {\it Gaia} (\url{https://www.cosmos.esa.int/gaia}), processed by the {\it Gaia} Data Processing and Analysis Consortium (DPAC,
\url{https://www.cosmos.esa.int/web/gaia/dpac/consortium}). Funding for the DPAC has been provided by national institutions, in particular the institutions participating in the {\it Gaia} Multilateral Agreement.

%

\vspace{5mm}

\facilities{ TESS, Gaia, Palomar: 5.1m, WIYN: 3.6m, LCOGT: 1m, LCOGT: 0.4m, OSN: 1.5m, GdP: 0.4m, CMO: 0.6m, CDK20: 0.5m, CDK14: 0.36m}

\software{ astropy \citep{2013A&A...558A..33A,2018AJ....156..123A}, AstroImageJ \citep{Collins2017}, juliet \citep{juliet}, batman \citep{Kreidberg2015}, radvel \citep{Fulton2018}}



\appendix

\section{List of known planet candidates missed by our detection pipeline}
Table \ref{missed_TOIs} shows a list of known planet candidates missed by our detection pipeline. Most of these candidates have low BLS signal-to-noise ratio. 

\begin{table*}
    \centering
    {\renewcommand{\arraystretch}{0.7}
    \caption{Known TOIs missed by our detection pipeline. Candidate information is retrieved from ExoFOP.}
    \begin{tabular}{lcccccccc}
        \hline\hline
        TIC       &TOI &Tmag   &Period (days) &$R_{p}\ (R_{\oplus})$ &$\rm SNR_{transit}$ &$\rm SNR\ Ratio$ &$\delta_{\rm odd}/|\delta_{\rm odd}-\delta_{\rm even}|^{[1]}$  &$\delta_{\rm even}/|\delta_{\rm odd}-\delta_{\rm even}|$\\\hline
        1133072 &566 &12.63 &0.85 &1.6 &9.9 &1.6 &- &-\\
        4070275 &4364 &11.40 &5.42 &2.1 &5.8 &1.1 &- &-\\
        11996814 &2022 &11.65 &0.45 &5.8 &16.8 &1.3 &- &-\\
        28900646 &1685 &11.11 &0.67 &1.5 &5.3 &1.1 &- &-\\
        32497972 &876 &11.53 &29.48 &2.8 &5.9 &1.1 &- &-\\
        54962195$^{*,[2]}$ &663 &11.76 &2.60 &2.3 &8.3 &1.5 &- &-\\
        55488511 &557 &11.64 &3.34 &2.3 &5.6 &0.9 &- &- \\
        59128183 &2453 &12.42 &4.44 &3.0 &5.4 &0.9 &- &-\\
        71347873 &2293 &11.81 &6.07 &2.2 &8.4 &1.0 &- &-\\
        104208182 &1738 &12.49 &3.70 &3.6 &7.4 &1.3 &- &-\\
        119081096 &716 &12.43 &0.84 &3.0 &10.6 &1.2 &- &-\\
        124235800 &4898 &11.93 &2.76 &3.5 &7.3 &7.4 &- &-\\
        138762614 &1802 &11.13 &16.80 &2.5 &5.6 &1.1 &- &-\\
        140687214 &4327 &12.13 &0.83 &3.4 &5.0 &0.9 &- &-\\
        141527579 &698 &12.13 &15.09 &2.1 &5.1 &0.8 &- &-\\
        147892178 &5207 &13.28 &24.69 &6.8 &8.9 &1.6 &-&-\\
        149788158 &727 &11.00 &4.72 &2.0 &5.1 &1.0 &- &-\\
        154616309 &3397 &13.32 &3.63 &6.8 &10.9 &1.9 &2.82 &3.82\\
        154940895 &4572 &11.92 &26.95 &1.8 &5.0 &1.0 &- &-\\
        168751223 &2331 &13.37 &4.72 &7.5 &7.4 &1.3 &- &-\\
        198211976 &2283 &11.21 &0.40 &0.6 &6.2 &1.0 &- &-\\
        200593988 &526 &12.31 &7.70 &6.2 &7.9 &1.4 &- &-\\
        201186294 &1634 &11.01 &0.99 &1.79 &6.0 &1.1 &- &-\\
        219175972$^{*}$ &2441 &12.83 &12.89 &2.9 &9.3 &1.7 &- &-\\
        219195044$^{*}$ &714 &11.54 &4.32 &1.5 &5.1 &1.1 &- &-\\
        219229644 &870 &10.78 &22.03 &2.3 &5.7 &1.1 &- &-\\
        219698776 &1243 &11.20 &4.66 &2.5 &6.8 &1.3 &- &-\\
        220459976 &285 &12.17 &32.33 &3.0 &5.5 &0.9 &- &-\\
        224298134$^{*}$ &2079 &10.85 &1.49 &1.7 &10.5 &1.0 &- &-\\
        233602827$^{*}$ &1749 &12.26 &4.49 &2.0 &8.0 &1.5 &- &-\\
        235678745$^{*}$ &2095 &11.08 &17.66 &1.4 &5.3 &0.9 &- &-\\
        236934937 &2291 &12.00 &9.41 &2.5 &6.1 &1.1 &- &-\\
        237920046 &873 &12.12 &5.93 &1.7 &6.2 &1.0 &- &-\\
        240968774 &1467 &10.60 &5.97 &1.8 &9.4 &1.2 &- &-\\
        244170332 &5530 &11.40 &0.48 &1.1 &5.5 &0.7 &- &-\\
        261257684$^{*}$ &904 &10.85 &10.88 &2.6 &7.7 &1.4 &- &-\\
        267561446 &1284 &12.53 &1.28 &2.6 &9.2 &1.9 &- &-\\
        270355392$^{*}$ &4643 &10.61 &5.03 &1.4 &5.6 &1.0 &- &-\\
        271596225$^{*}$ &797 &11.71 &1.80 &1.3 &6.4 &0.9 &- &-\\
        274662200 &1285 &10.93 &1.23 &1.9 &6.2 &1.2 &- &-\\
        277833995 &5524 &11.76 &2.30 &2.0 &4.7 &1.0 &- &-\\
        284441182 &1470 &11.48 &2.53 &2.2 &14.4 &1.3 &- &-\\
        287139872$^{*}$ &1752 &12.75 &0.94 &2.0 &7.2 &1.3 &- &-\\
        298428237 &4574 &11.75 &0.77 &1.6 &4.8 &1.0 &- &-\\
        307849973$^{*}$ &4567 &11.92 &0.84 &1.4 &5.2 &0.9 &- &-\\
        318836983 &5532 &11.52 &5.65 &2.2 &5.7 &0.5 &- &-\\
        321669174 &2081 &11.64 &10.51 &1.8 &6.5 &1.1 &- &-\\
        322270620 &1083 &12.09 &12.98 &3.2 &12.0 &2.2 &0 &1\\
        329148988 &2285 &11.31 &27.27 &1.9 &5.0 &0.9 &- &-\\
        332477926 &1754 &11.72 &16.22 &2.5 &6.8 &0.9 &- &-\\
        348673213 &1639 &12.97 &0.90 &3.4 &7.9 &1.4 &- &-\\
        348755728 &1883 &13.35 &4.51 &5.9 &10.3 &2.2 &1.91 &2.91\\
        351601843 &1075 &11.12 &0.60 &1.9 &6.6 &1.1 &- &-\\
        353475866 &1693 &10.67 &1.77 &1.4 &4.8 &0.8 &- &-\\
        354944123$^{*}$ &4342 &11.03 &5.54 &2.3 &8.9 &1.6 &- &-\\
        359357695 &1880 &13.06 &1.73 &6.0 &9.4 &1.3 &- &-\\ 
        364074068 &1756 &12.09 &2.78 &1.8 &13.3 &1.7 &0.27 &1.27\\
        374829238 &785 &11.50 &18.63 &1.2 &12.6 &1.3 &- &-\\
        389371332 &4346 &12.44 &3.91 &1.6 &5.7 &1.0 &- &-\\
        422756130 &1695 &11.03 &3.13 &1.8 &6.0 &1.0 &- &-\\
        424747720 &4188 &13.01 &10.28 &11.87 &17.9 &3.3 &0.03 &1.03\\
        441738827$^{*}$ &2084 &13.33 &6.08 &2.6 &8.0 &1.5 &- &-\\
        441739871 &1763 &12.81 &3.80 &1.9 &5.7 &1.1 &- &-\\
        441798995$^{*}$ &2269 &11.95 &2.84 &1.5 &5.8 &1.2 &- &-\\
        458419328 &3785 &12.50 &4.67 &4.9 &15.3 &3.1 &3.04 &2.04\\
        468777766 &3750 &12.99 &12.48 &8.5 &16.2 &2.6 &0 &1\\
        470987100 &1732 &11.33 &4.12 &2.6 &8.6 &1.7 &- &-\\
        \hline\hline 
    \end{tabular}
    \begin{tablenotes}
      \item[1]  [1]\ We only calculate the depth consistency if SNR$\geq 10$ and SNR Ratio$\geq 1.5$. 
      \item[2]  [2]\ Targets marked with ``*'' are systems with multi candidates. Here we list the planet candidate with highest SNR reported by \tess\ team.
    \end{tablenotes}
    \label{missed_TOIs}}
\end{table*}

\section{Vetting Plots}

\begin{figure*}
\includegraphics[scale=1]{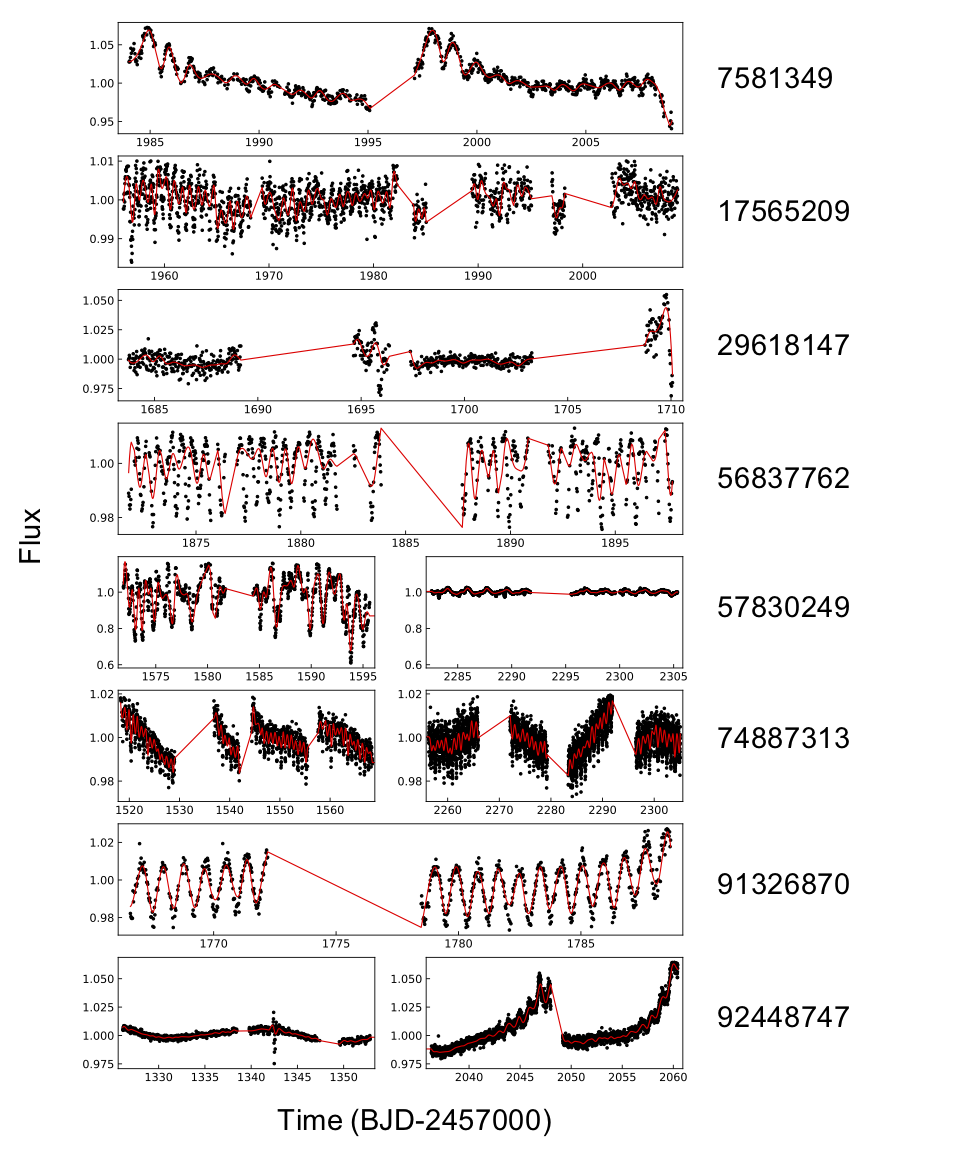}
\caption{Light curves of 44 false positives removed through visual inspection in Section \ref{photometry_check}. The red solid line is our spline model used to detrend the data. The light curves of targets that have data from the TESS extended mission are shown in two panels. The target name (TIC) is listed on the right.}
\label{detrending_issues}
\end{figure*}

Figure \ref{detrending_issues} shows the light curves of 44 false positives removed through visual inspection in Section \ref{photometry_check}. Figure \ref{odd_even_secondary} shows an example diagnostic plot of the odd/even and secondary analysis for a planet candidate around TIC 224283851 alerted by our detection pipeline. Figure \ref{phase_variation} shows an example diagnostic plot of the synchronization analysis for a planet candidate around TIC 329884233 alerted by our detection pipeline.

\begin{figure*}
\includegraphics[scale=1]{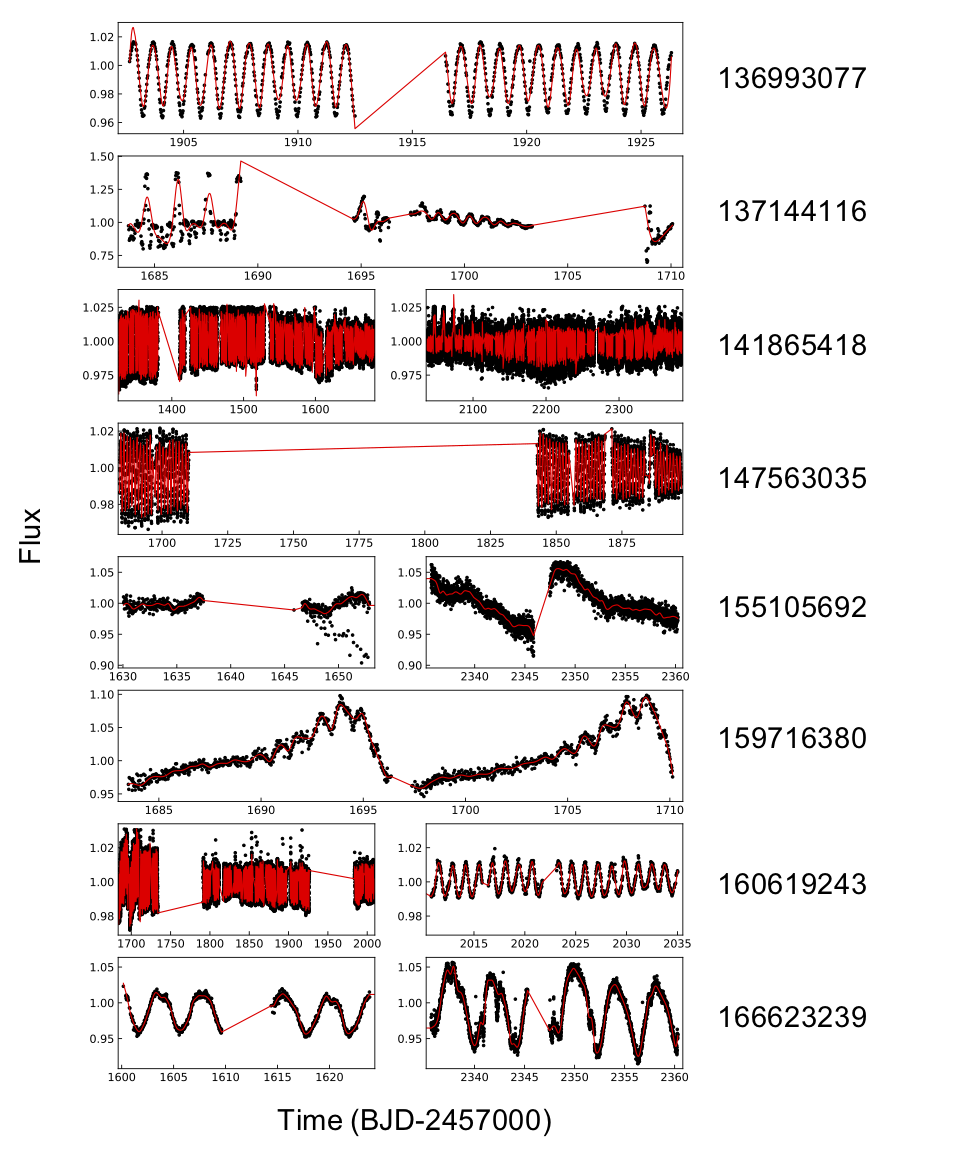}
\label{tab:continued}
\end{figure*}

\begin{figure*}
\includegraphics[scale=1]{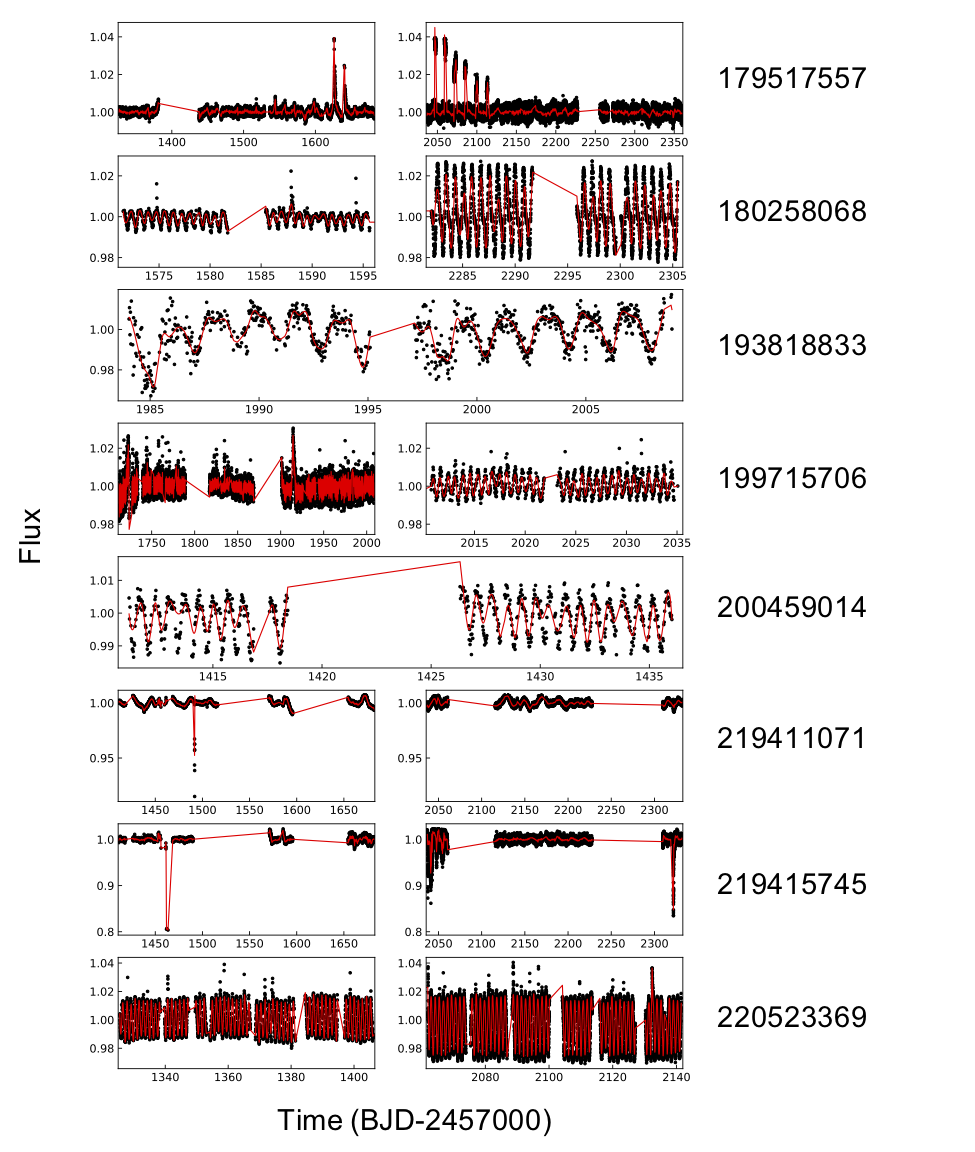}
\label{tab:continued}
\end{figure*}

\begin{figure*}
\includegraphics[scale=1]{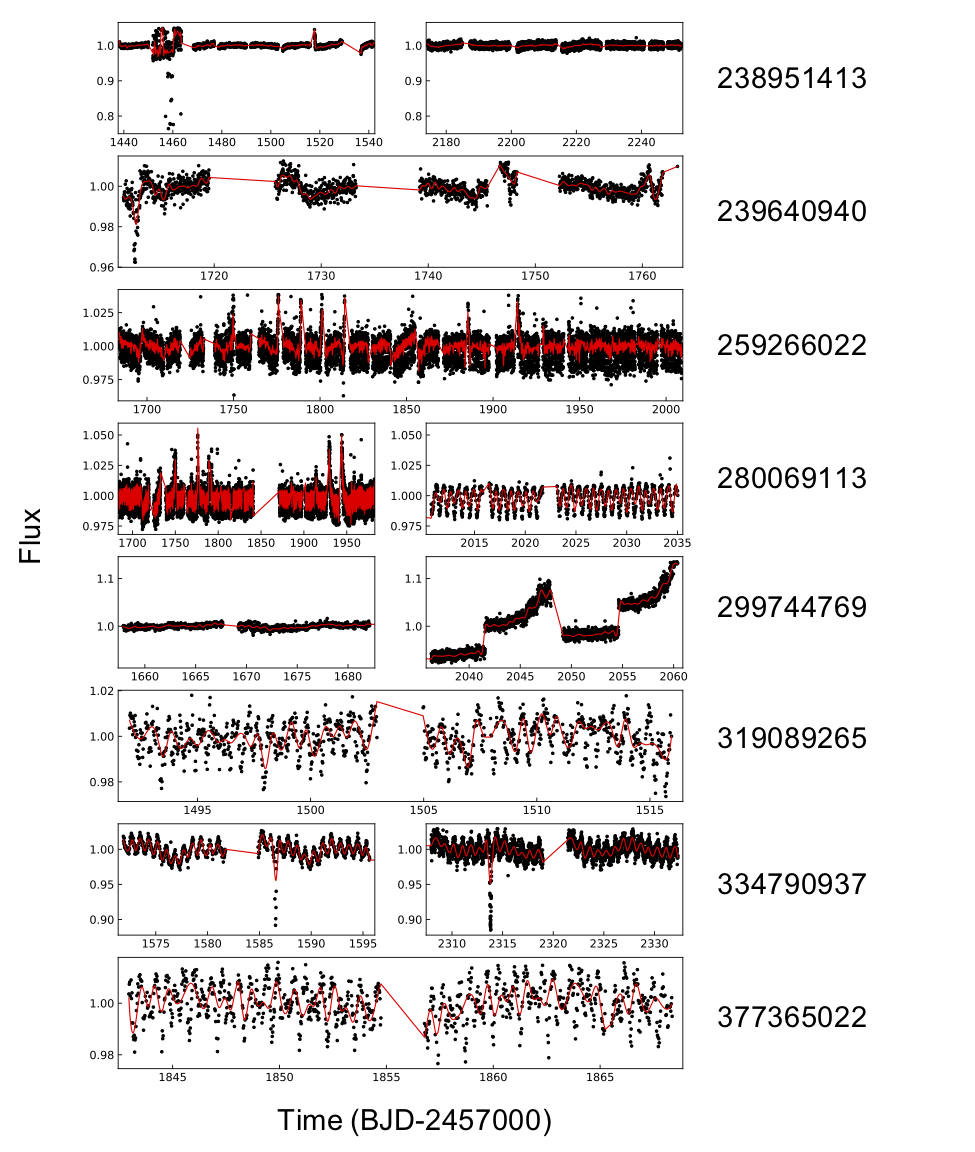}
\label{tab:continued}
\end{figure*}

\begin{figure*}
\includegraphics[scale=1]{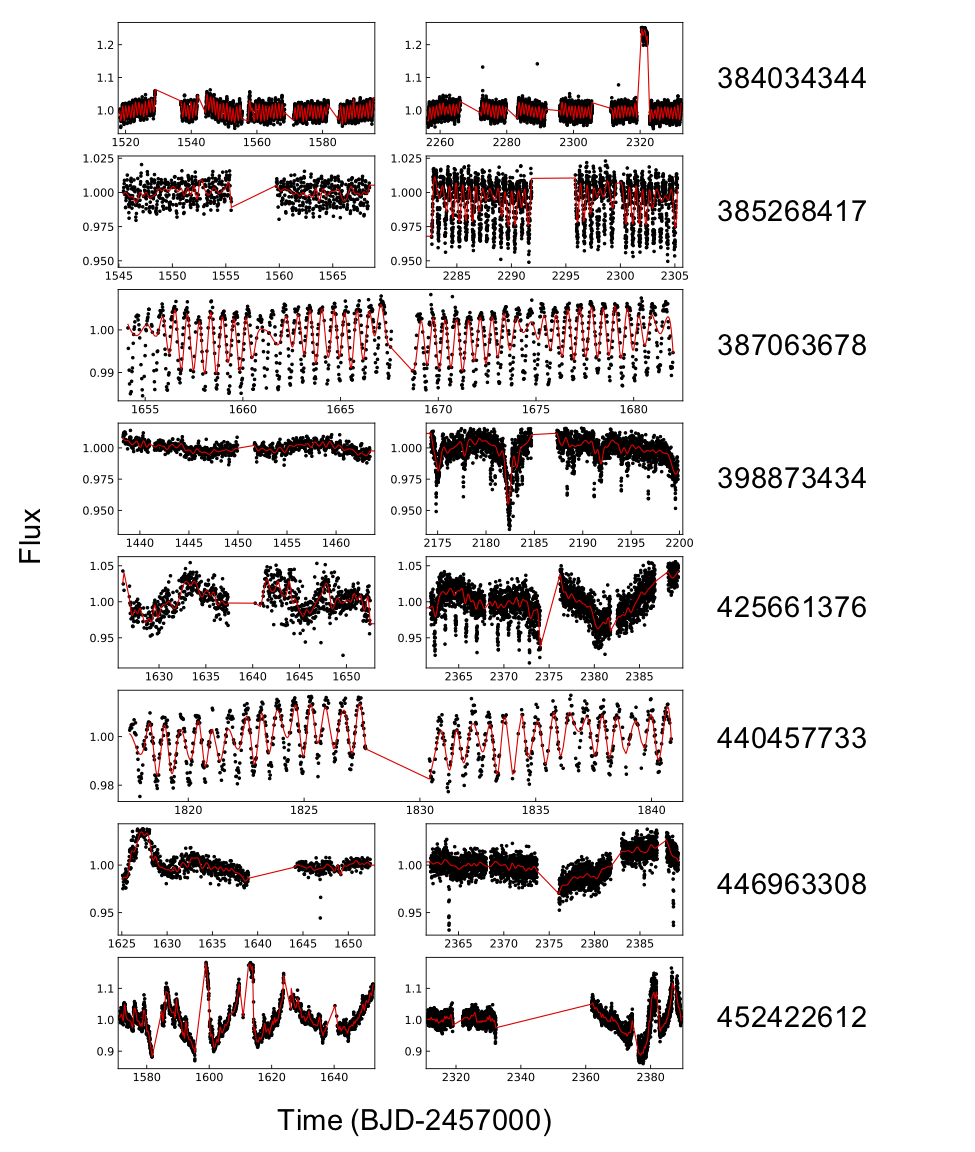}
\label{tab:continued}
\end{figure*}

\begin{figure*}
\includegraphics[scale=1]{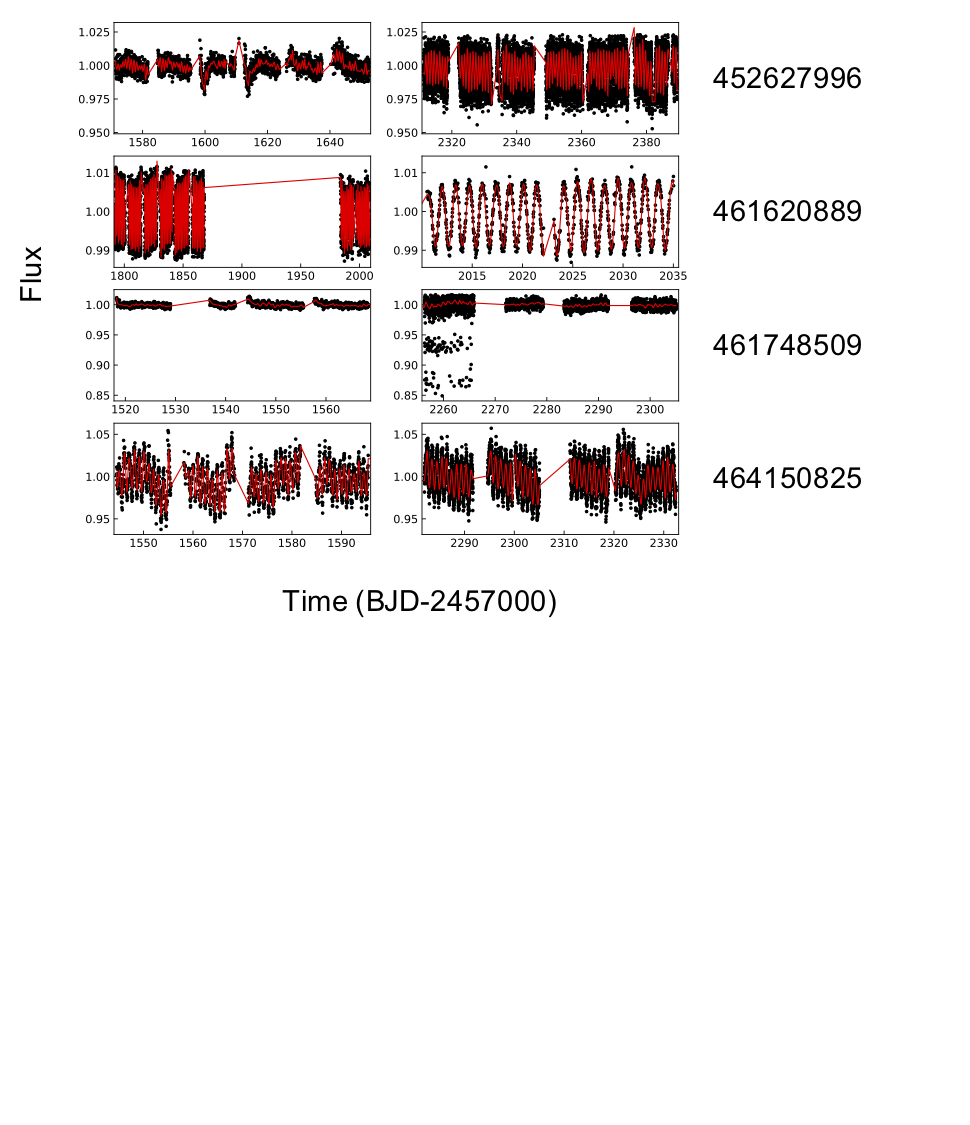}
\label{tab:continued}
\end{figure*}

\begin{figure*}
\centering
\includegraphics[width=0.7\textwidth]{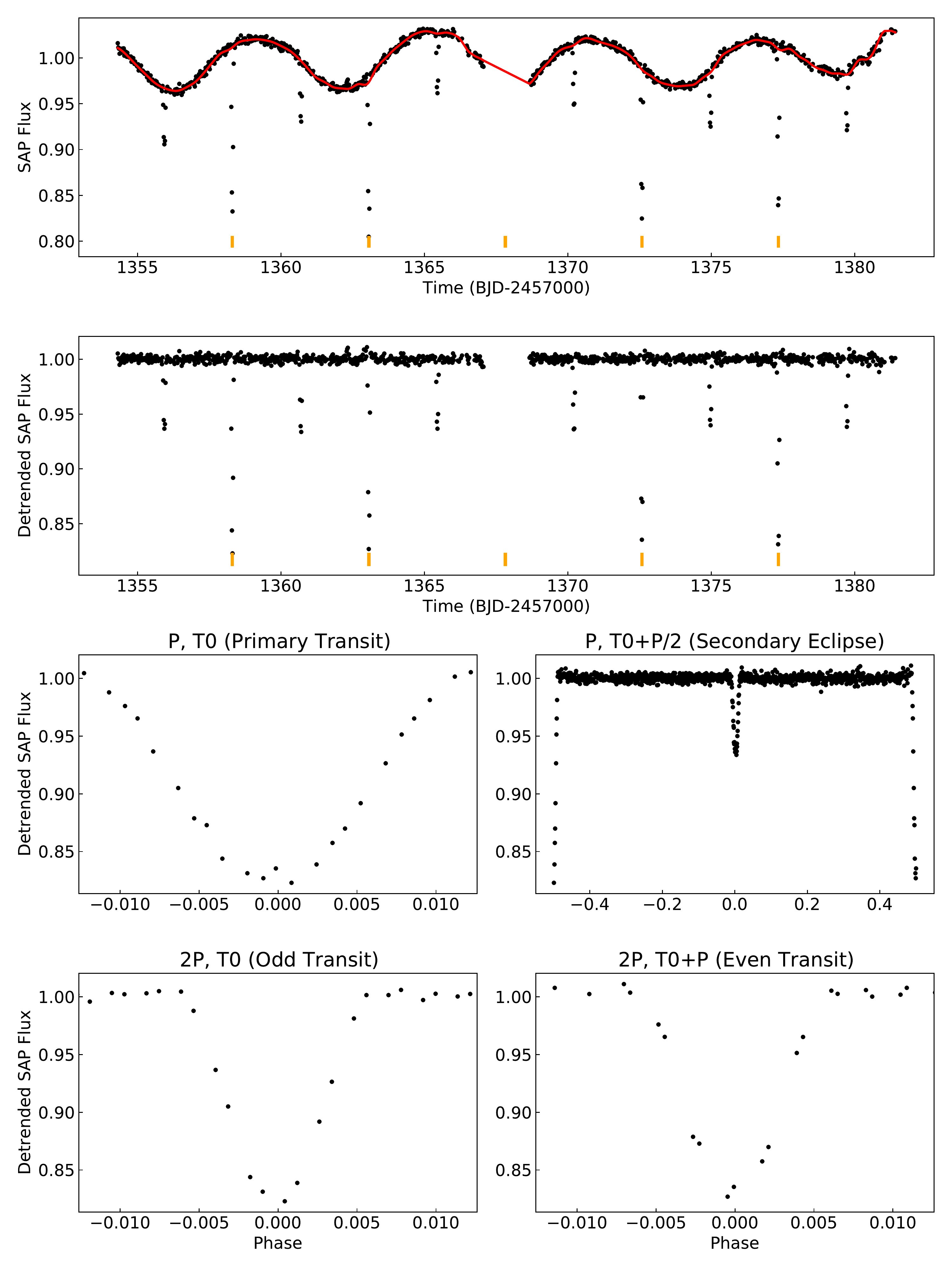}
\caption{An example diagnostic plot of the odd/even and secondary analysis for TIC 224283851. {\it Top row}: the raw QLP light curve. The cubic spline model (binning size=0.3 days) used for detrending is shown as a red solid curve. {\it Second row}: the detrended QLP light curve. The orange ticks in these panels mark the signal of primary transit.} {\it Third row, left panel}: phase-folded light curves at the best period and mid-transit time found by the detection pipeline (primary transit). {\it Third row, right panel}: phase-folded light curves at best period but shift a half period (secondary eclipse). {\it Bottom row, left panel}: phase-folded light curves at twice of the best period and mid-transit time found by the detection pipeline (odd transit). {\it Bottom row, right panel}: phase-folded light curves at twice of the best period but shift a period (even transit).
\label{odd_even_secondary}
\end{figure*}

\begin{figure*}
\includegraphics[width=0.99\textwidth]{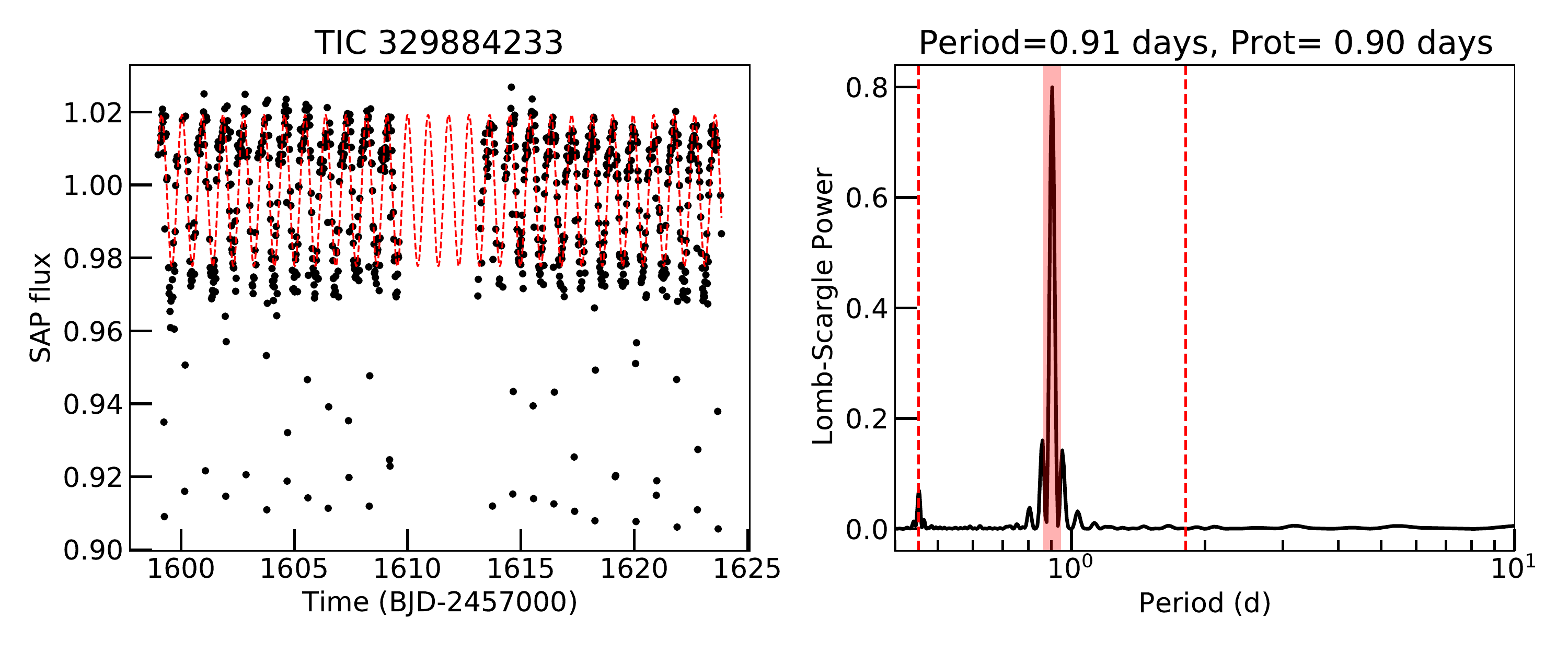}
\caption{An example diagnostic plot of the synchronization test for TIC 329884233. {\it Left panel}: The raw QLP light curve of the target star. The red dashed curve is the best Lomb-Scargle model fit of the light curve after masking out the transit signals based on the period and mid-transit time found by the detection pipeline. {\it Right panel}: The corresponding Lomb-Scargle periodogram. The rotation period is marked as a red shade region. Two nearby aliasing periods are marked by red vertical dashed lines. The measured transiting and rotation periods are listed on the top of the plot. }
\label{phase_variation}
\end{figure*}

\section{Prior setting on the light curve modeling}
Table \ref{transit_fit_prior} shows the prior settings for the \tess\ light curve modeling of each planetary system (see Section \ref{lightcurve_modeling}).
\begin{table*}
    \centering
    \caption{Prior settings for the \tess\ light curve modeling of each planet candidate.}
    \begin{tabular}{lcr}
        \hline\hline
        Parameter       &Prior     &Description\\\hline
        \it{Planet candidate parameters}\\
        $P_b$ (days)  &$\mathcal{N}^{[1]}$ ($P^{[2]}_{\rm BLS}$\ ,\ $0.2^{2}$)
        &Orbital period of companion in the system.\\
        $T_{0,b}$ (BJD-2457000)    &$\mathcal{N}$ ($T_{\rm 0,BLS}$\ ,\ $0.2^{2}$) 
        &Mid-transit time of the companion in the system.\\
        $r_{1}$    &$\mathcal{U}^{[3]}$ (0\ ,\ 1)
        &Parametrisation for {\it p} and {\it b} of the companion in the system.\\
        $r_{2}$    &$\mathcal{U}$ (0\ ,\ 1)
        &Parametrisation for {\it p} and {\it b} of the companion in the system.\\
        $e_{b}$   &$0$ (Fixed)  &Orbital eccentricity of the companion in the system.\\
        $\omega_{b}$ (deg)  &$90$ (Fixed)  &Argument of periapsis of the companion in the system.\\
        \\
        \it{\tess\ photometry parameters}\\
        $D_{\rm TESS}$  &$\mathcal{TN}^{[4]}$ ($D^{[5]}$\ ,\ $0.05^{2}$,\ 0\ ,\ 1)      &\tess\ photometric dilution factor.\\
        $M_{\rm TESS}$ (ppm)    &$\mathcal{N}$ (0\ ,\ $0.1^{2}$)      &Mean out-of-transit flux of \tess\ photometry.\\
        $\sigma_{\rm TESS}$ (ppm) &$\mathcal{J}^{[6]}$ ($10^{-6}$\ ,\ $10^{6}$)      &\tess\ additive photometric jitter term.\\
        $q_{1}$        &$\mathcal{U}$ (0\ ,\ 1)  &Quadratic limb darkening coefficient.\\
        $q_{2}$        &$\mathcal{U}$ (0\ ,\ 1)  &Quadratic limb darkening coefficient.\\
        \\
        \it{Stellar parameters}\\
        ${\rho}_{\ast}$ ($\rm kg\ m^{-3}$)  &$\mathcal{N}$ ($\rho_{\ast}$\ ,\ $\sigma_{\rho_{\ast}}^{2}$) &Stellar density.\\
        \hline\hline 
    \end{tabular}
    \begin{tablenotes}
    \item[1]  [1]\ $\mathcal{N}$($\mu,\ \sigma^{2}$) means a normal prior with mean $\mu$ and standard deviation $\sigma$. 
    \item[2]  [2]\ The priors of orbital period and mid-transit time are centered at the values found by the high resolution BLS search.
    \item[3]  [3]\ $\mathcal{U}$($a,\ b$) stands for a uniform prior ranging from a to b.
    \item[4]  [4]\ $\mathcal{TN}$($\mu,\ \sigma^{2},\ a\ , b$) stands for a truncated normal prior with mean $\mu$ and standard deviation $\sigma$ ranging from $a$ to $b$.
    \item[5]  [5]\ We use the light contamination ratio from \tess\ Input Catalog (TIC) v8 \citep{Stassun2019tic} and transform it into the dilution factor $D$ (see Section \ref{sample_selection}). 
    \item[6]  [6]\ $\mathcal{J}$($a,\ b$) stands for a Jeffrey's prior ranging from a to b.
    \end{tablenotes}
    \label{transit_fit_prior}
\end{table*}


\section{Candidates removed in the light curve modeling section.}
\begin{figure*}
\includegraphics[width=0.99\textwidth]{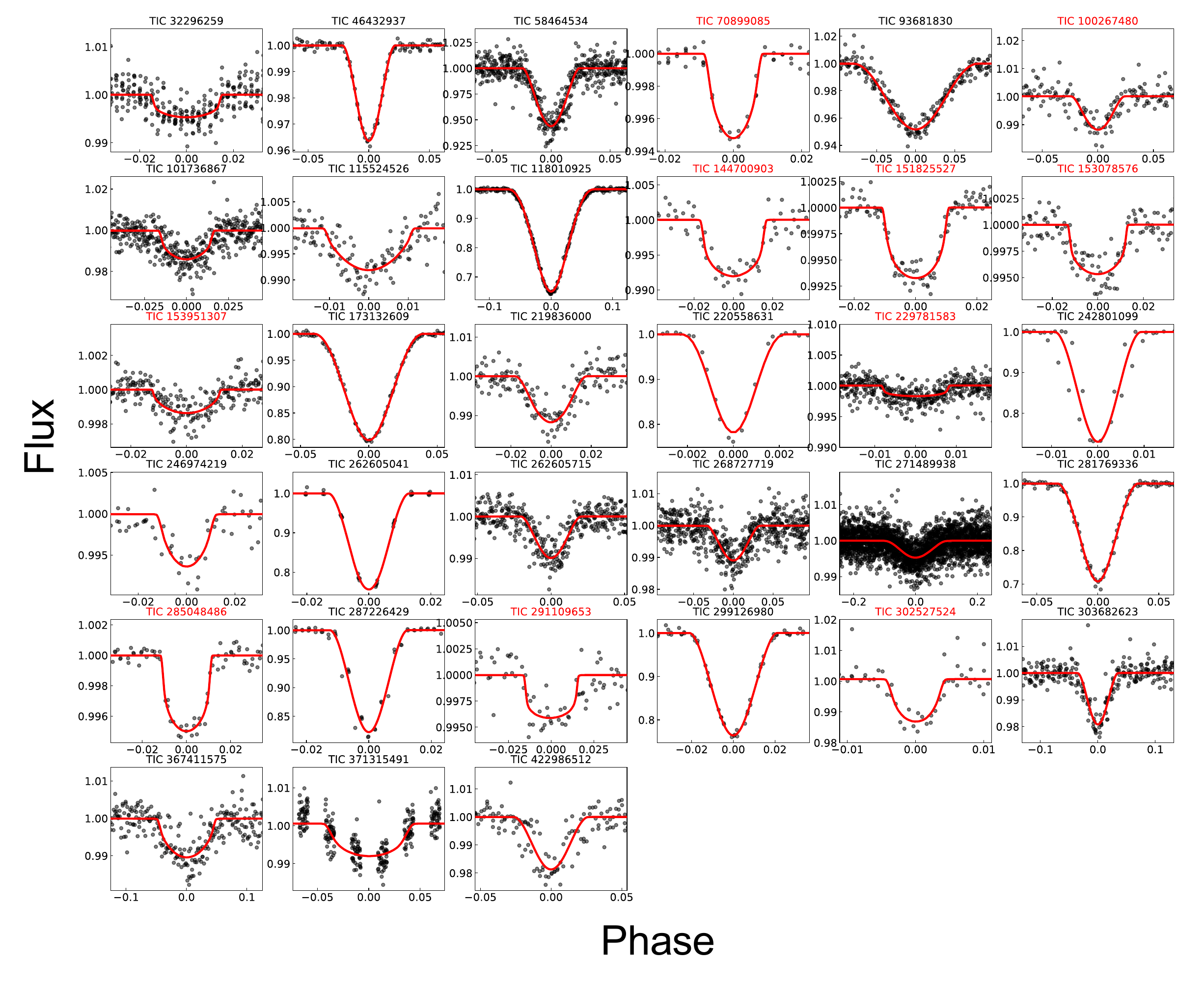}
\caption{Phase-folded light curves of 33 candidates removed in the modeling step along with the best fit transit models. The target name is shown at the top of each panel. The titles of all TOIs are marked in red. Details of these targets are listed in Table \ref{removed_candidates_from_modeling}.}
\label{removed_candidates_from_modeling_lc}
\end{figure*}

Table \ref{removed_candidates_from_modeling} shows the list of candidates removed in the light curve modeling section step (see Section \ref{lightcurve_modeling}) that are outside our selection function in terms of radius, impact parameter or orbital period. We show their light curves along with the best-fit transit models in Figure \ref{removed_candidates_from_modeling_lc}.

\begin{table*}
    \centering
    \caption{List of 33 candidates removed from our analysis in the light curve modeling step that are outside our selection function in terms of (1) radius range ($7\ R_{\oplus}\leq R_{p}\leq 2\ R_{J}$), (2) impact parameter range ($b\leq0.9$)  or (3) period range ($0.8 \leq P_{b}\leq 10$ days).}
    \begin{tabular}{lcccccccc}
        \hline\hline
        TIC       &TOI    &Period (days) &Impact parameter b &$R_{p}\ (R_{J})$ &Comment \\\hline
        32296259 &-	&2.778 &0.21 &0.42 &1\\
        46432937 &-	&1.437 &1.56 &4.71 &1, 2 \\
        58464534 &- &1.403 &1.57 &5.10 &1, 2\\
        70899085 &442 &4.052 &0.71 &0.44 &1\\
        93681830 &- &1.848 &1.53 &5.46 &1, 2\\
        100267480 &2341	&0.877 &1.61 &4.76 &1, 2\\
        101736867 &- &2.648 &0.16 &0.59 &1\\
        115524526 &- &4.657 &0.12 &0.45 &1\\
        118010925 &- &0.729 &0.42 &3.88 &1, 3\\
        144700903 &532 &2.326 &0.20 &0.51 &1\\
        151825527 &672 &3.634 &0.33 &0.41 &1\\
        153078576 &2407 &2.703 &0.17 &0.32 &1\\
        153951307 &1238	&3.295 &0.26 &0.20 &1\\
        173132609 &- &1.079 &0.68	&3.11 &1\\
        219836000 &- &1.585 &1.62	&4.99 &1, 2\\
        220558631$^{[1]}$ &- &5.180 &1.48	&4.53 &1, 2\\
        229781583 &1245 &4.820 &0.36 &0.20 &1\\
        242801099$^{[2]}$ &- &9.135 &0.97 &4.37 &1, 2\\
        246974219 &- &1.909 &0.72 &0.42 &1\\
        262605041 &- &3.666 &1.26 &4.51 &1, 2\\
        262605715 &- &1.161 &1.74 &5.26 &1, 2\\
        268727719 &- &0.626 &1.76 &5.43 &1, 2 ,3\\
        271489938 &- &0.489 &1.82 &5.61 &1, 2, 3\\
        281769336 &- &1.925 &0.6 &4.63 &1\\
        285048486 &1728	&3.491 &0.46 &0.41 &1\\
        287226429 &- &4.927 &1.26	&4.55 &1, 2\\
        291109653 &5486 &2.025 &0.31 &0.33 &1 \\
        299126980 &- &3.287 &1.09 &5.88 &1, 2\\
        302527524 &2952	&10.784 &0.67 &0.59 &1, 3\\
        303682623 &- &0.679 &1.71 &5.24 &1, 2, 3\\
        367411575 &- &1.193 &0.13 &0.57 &1\\
        371315491 &- &0.406 &0.66 &0.55 &1, 3\\
        422986512 &- &1.115 &1.66 &5.18 &1, 2\\
        \hline\hline 
    \end{tabular}
    \begin{tablenotes}
       \item[1]  [1]\ The real period of this system is 36 days. Two signals separated by 5.18 days are the primary and secondary of the eccentric eclipsing binary with similar depth.
       \item[2]  [2]\ Two signals separated by 9.13 days are probably the primary and secondary of a long-period eclipsing binary.
    \end{tablenotes}
    \label{removed_candidates_from_modeling}
\end{table*}

\section{Details of all ground follow-up observations for each candidate}\label{detail_ground_obs}

We present the details of ground-based follow-up observations for five candidates in Table \ref{final_candidates} used for statistics in this work. 

\subsection{TIC 20182780 (TOI-3984)}\label{TOI3984}
We obtained two focused ground-based follow-up observations for TIC 20182780 (TOI-3984) using the 1m Las Cumbres Observatory Global Telescopes \citep[LCOGT;][]{Brown2013} on 2022 April 14 and June 6 in the $i'$ and $g'$ bands with an exposure time of 150 and 300s. The Sinistro cameras have a $26' \times 26'$ field of view (FOV) as well as a plate scale of $\rm 0.389''$ per pixel. The images of both observations have stellar point-spread-functions (PSF) with a full-width-half-maximum (FWHM) of $2.2\arcsec$ and $2.3\arcsec$, respectively. We also acquired focused alternating $V\&I$ band observations using the 1.5m telescope at Observatorio de Sierra Nevada on 2022 May 10, which has a FOV of $18'\times13'$ and a pixel scale of $0.455\arcsec$. The exposure times we set are 100s and 50s for V and I band observations. The estimated PSF is about $2.7\arcsec$.

In addition, we took an AO observation for TIC 20182780 using the Palomar High Angular Resolution Observer (PHARO) on the Palomar-5.1m telescope on 2022 February 13 in Br-$\gamma$ band ($\lambda_{o}=2.2$ $\mu$m) to search for stellar companions. The result reveals that it is an isolated star, with no companions 6.1 magnitudes fainter than the target out to 0.5$\arcsec$.

To constrain the companion mass, we also obtain seven RVs with the WIYN/NEID spectrograph \citep{Schwab2016} for it between 2022 March 13 and 2022 April 9 with a baseline of 27 days. The observations are queue scheduled. Each exposure took 1200s and the median RV precision is about 25 m/s. The NEID RVs put a $3\sigma$ mass upper limit of $0.32\ M_{J}$ on the companion mass and rule out the stellar binary scenario. We refer the readers to \cite{Wang2022} for more information about NEID observation and data reduction. The final RV data are listed in Table \ref{neidrv}.

\begin{table*}
     \centering
     \caption{Seven NEID RV measurements of TIC 20182780 (TOI-3984). Each observation took an exposure time of 1200s.}
     \begin{tabular}{ccc}
         \hline\hline
         BJD$_\mathrm{TDB}$       &RV\ (m~s$^{-1}$) &$\sigma_{\rm RV}$\ (m~s$^{-1}$) \\\hline
        2459651.890485 &-5463.1 &25.8\\
        2459656.839133 &-5385.9 &22.2\\
        2459657.761302 &-5423.9 &37.0\\
        2459663.895137 &-5415.4 &17.1\\
        2459664.861012 &-5379.7 &23.9\\
        2459671.889475 &-5387.5 &27.6\\
        2459678.892503 &-5524.9 &21.0\\
          \hline
     \end{tabular}
     \label{neidrv}
\end{table*}

\subsection{TIC 71268730 (TOI-5375)}
We collected two ground focused photometric observations for TIC 71268730 (TOI-5375) using the GdP-0.4m and CMO-0.6m at Grand-Pra and Caucasian Mountain Observatory. The GdP-0.4m has a FOV of $12.9' \times 12.5'$ and a pixel scale of $0.73\arcsec$. The observation was taken on 2022 March 5 in a clear filter with an exposure time of 180s. The seeing is good with a light curve RMS of 0.0074. The CMO-0.6m has a FOV of $22' \times 22'$ and a pixel scale of $0.67\arcsec$ \citep{Berdnikov2020}. The observation was taken on 2022 March 31 in $R_{c}$ band with an exposure time of 120s. The PSF of two observations are $4.5\arcsec$ and $2.4\arcsec$. We used 10 and 7 comparison stars with an aperture of $6.6\arcsec$ and $4.7\arcsec$ to do the photometric analysis for these two observations, respectively.

\subsection{TIC 79920467 (TOI-3288)}
We collected three LCOGT light curves for TIC 79920467 (TOI-3288), one of them was done with the 0.4m while two were done with the 1m. LCOGT-0.4m has a FOV of $29' \times 19'$ with a pixel scale of $0.571\arcsec$. The LCOGT-0.4m observation was carried out in the $i'$ band on 2021 June 7 with an exposure time of 200s. The LCOGT-1m observations were carried out in the $i'$ and $g'$ bands on 2021 June 19 and 2022 May 16 with an exposure time of 100s and 300s. The images are all focused with a PSF of $2.6\arcsec$, $2.3\arcsec$, and $2.0\arcsec$, respectively. We used an aperture of $5.1\arcsec$ to reduce the LCOGT-0.4m data while $1.6\arcsec$ for the LCOGT-1m data. We also obtained two luminous band observations for this target using a CDK20-0.5m at El Sauce Observatory, Chile on 2021 September 2 and 2021 October 28, under a good seeing condition. For these observations, the exposure time was set at 120s. The CDK20-0.5m has a FOV of $35.87' \times 35.87'$ and a pixel scale of $0.52\arcsec$.

\subsection{TIC 95057860 (TOI-4201)}
We acquired a total of five focused LCOGT-1m observations for TIC 95057860 (TOI-4201). The first observation was done in the $i'$ band on 2021 September 1. Two $g'$ and $i'$ band alternating observations (four light curves) were carried out on 2021 September 26 and 2021 October 13. All $i'$ band observations were taken with an exposure time of 180s while $g'$ band observations have a 300s exposure time. The photometric apertures we used for three observations are $8.5\arcsec$, $8.5\arcsec$ and $6.2\arcsec$. The PSFs of three observations are $2.3\arcsec$, $5.1\arcsec$, and $3.3\arcsec$, respectively.

\subsection{TIC 382602147 (TOI-2384)}
We collected two ground-based follow-up light curves for TIC 382602147 (TOI-2384). The first observation was obtained with the Evans telescope at El Sauce Observatory, Chile, a 0.36m Corrected Dall Kirkham, in $R_{c}$ band on 2020 November 9. The telescope was fitted with an SBIG 1603-3 CCD with 1536x1024 pixels binned 2x2 in camera for an image scale of 1.47$\arcsec$/pixel, giving a field of view of $18.8\arcmin \times 12.5\arcmin$. The calibrated data consisting of 105 exposures of 180 seconds was analysed with a circular aperture of 5.9" radius in AstroImageJ. Another observation was done with LCOGT-1m in $g'$ on 2021 August 5. We reduced the data with a $4.7\arcsec$-size circular aperture. The PSFs of two observations are $3.3\arcsec$ and $2.4\arcsec$, respectively.



\bibliography{planet}{}
\bibliographystyle{aasjournal}



\end{document}